\documentclass[twocolumn,secnumarabic,amssymb, nobibnotes, aps, prb,groupedaddress,superscriptaddress]{revtex4-2}
\usepackage{amsmath,graphicx,latexsym,times,color}
\usepackage{setspace}
\usepackage{hyperref}
\usepackage{array}
\usepackage{textcomp}
\usepackage{titlesec}
\usepackage{physics}
\usepackage{gensymb}
\usepackage{nccmath}
\usepackage{soul} 
\usepackage{empheq} 

\usepackage{ulem} 
\usepackage[dvipsnames]{xcolor}
\definecolor{blue-violet}{rgb}{0.54, 0.17, 0.89}\newcommand{\V}[1]{\ensuremath{\mathbf{#1}}} 

\let\oldtimes\times  
\renewcommand\times{{\oldtimes}}

\definecolor{darkorchid}{HTML}{bf3eff}

\begin{document}

\title{Extended saddle points govern long-lived antiskyrmions}

	\author{Megha Arya}
	\affiliation{Universit\'e de Toulouse, CNRS, CEMES, Toulouse, France}

    \author{Moritz A. Goerzen}
	\affiliation{Universit\'e de Toulouse, CNRS, CEMES, Toulouse, France}

  	\author{Lionel Calmels}
	\affiliation{Universit\'e de Toulouse, CNRS, CEMES, Toulouse, France}

  	\author{Shiwei Zhu}
	\affiliation{Universit\'e de Toulouse, CNRS, CEMES, Toulouse, France}

  	\author{Bhanu Jai Singh}
	\affiliation{Universit\'e de Toulouse, CNRS, CEMES, Toulouse, France}

	\author{Stefan Heinze}
	\affiliation{Institute of Theoretical Physics and Astrophysics, University of Kiel, Leibnizstrasse 15, 24098 Kiel, Germany}
	\affiliation{Kiel Nano, Surface, and Interface Science (KiNSIS), University of Kiel, 24118 Kiel, Germany}

 	\author{Dongzhe Li}
    \email[Contact author: ]{dongzhe.li@cemes.fr}
	\affiliation{Universit\'e de Toulouse, CNRS, CEMES, Toulouse, France}
	
	\date{\today}
	
	\begin{abstract}
Achieving long-lived nanoscale magnetic solitons remains a central challenge, as their lifetimes typically decrease rapidly with temperature. Here, we demonstrate that anisotropic Dzyaloshinskii–Moriya interaction (aDMI) enables spatially extended saddle points (SPs) that fundamentally alter thermally activated decay. 
In contrast to conventional localized SPs, these extended configurations completely suppress the entropic contribution to the activation rate, rendering the lifetimes effectively temperature independent.
To establish this mechanism, we develop a first-principles method based on spin spirals to compute DMI beyond the isotropic approximation, resolving its full directional dependence for arbitrary nearest neighbors. We apply this method to oxidized Fe$_3$GeTe$_2$ (FGT-O), an experimentally accessible van der Waals magnet. Oxygen adsorption simultaneously breaks inversion symmetry and lowers the in-plane crystalline symmetry, thereby generating a sizable aDMI. We demonstrate that aDMI stabilizes nanoscale antiskyrmions with energy barriers exceeding 120 meV at low external magnetic fields. Crucially, extended SPs enhance the lifetime in FGT-O by more than five orders of magnitude at room temperature compared to conventional ultrathin-film skyrmion systems. We further show that aDMI is not the only route to such extended SPs and identify the general conditions under which they emerge, establishing a general route to soliton decay pathways with temperature-independent prefactors. Our results uncover a new paradigm for enhancing soliton stability through transition-state geometry rather than energy-barrier height.


\emph{}\\
\emph{}

\noindent DOI: $\times \times \times$ \hfill Subject Areas: Condensed Matter Physics, Magnetism, Spintronics
	\end{abstract}
	
	\maketitle

\section{INTRODUCTION}

Topological spin textures -- such as magnetic skyrmions and antiskyrmions -- are currently of great interest in condensed matter physics due to their rich fundamental properties and potential applications in information technology \cite{fert2017magnetic, luo2018reconfigurable, gobel2021beyond, psaroudaki2021skyrmion, kim2019dynamics, moon2019existence}. Their existence results from a fine balance among different magnetic interactions, giving rise to chiral spin configurations. A key ingredient enabling such textures is the Dzyaloshinskii-Moriya interaction (DMI) \cite{yang2023first,Kuepferling2023}, which originates from spin-orbit coupling (SOC) in the presence of broken inversion symmetry and favors canted magnetization alignments.

In contrast to isotropic DMI (iDMI), which is currently under intense scrutiny both theoretically and experimentally \cite{Yang2015,Belabbes2016}, its anisotropic counterpart, known as anisotropic DMI (aDMI), is far less developed than iDMI but is a subject on the rise. So far, the study of aDMI has been largely restricted to square lattices \cite{hoffmann2017antiskyrmions,cui2022anisotropic,ga2022anisotropic}, whereas triangular lattices, which naturally host multiple competing symmetry directions and geometric frustration at the nearest-neighbor (NN) level, have received considerably less attention. Triangular lattices are also more commonly realized in real materials, further underscoring the importance of extending aDMI studies beyond square lattices.

The practical applications of magnetic solitons impose stringent requirements, including robust thermal stability near room temperature (RT), efficient and reliable control by external stimuli, and nanoscale dimensions, ideally below 5-10 nm. In particular, at nanoscale dimensions, thermal stability becomes a critical issue, since thermal fluctuations can trigger soliton collapse and result in information loss. The resulting lifetime ($\tau$) of topological solitons is governed by thermally activated processes and is commonly described by an Arrhenius law \cite{bessarab2012harmonic}:
\begin{equation}
\label{tau_Arrhenius}
\tau = \Gamma_{0}^{-1} \exp\left(\frac{\Delta E}{k_{\mathrm{B}} T}\right)~,
\end{equation}
where $\Delta E$, $\Gamma_{0}$, $k_{\mathrm{B}}$, and $T$ denote the energy barrier, pre-exponential factor, Boltzmann constant, and temperature, respectively. Here, we would like to emphasize that lifetime calculations are fundamentally challenging as they require navigating complex energy surfaces to identify the minimum energy path (MEP) for transition between states and necessitate determining both the energy barrier $\Delta E$ and the entropic contributions incorporated in the pre-exponential factor $\Gamma_0$ \cite{bessarab2012harmonic, wild2017entropy, desplat2018thermal, goerzen2022atomistic}.

To date, iDMI, exhibited by most of the studied systems, has been primarily associated with rotationally symmetric skyrmions and their collapse pathways mediated by compact (\textit{i.e.}, strongly localized) saddle points (SPs) \cite{bessarab2012harmonic,li2022strain,Dongzhe2024,arya2025new}. By contrast, aDMI reflects reduced in-plane symmetry and exhibits a crystallographic direction dependence in both its magnitude and orientation, acting as a symmetry-selective interaction rather than as a mere quantitative modification of iDMI. Although aDMI is known to favor antiskyrmionic textures \cite{hoffmann2017antiskyrmions,Xin2024,niu2025magnetic}, its role in thermally activated dynamics and lifetime has not yet been explored.

In the literature, strategies to enhance the lifetime of magnetic solitons have mainly focused on increasing the energy barrier $\Delta E$, because it enters exponentially in Eq.~(\ref{tau_Arrhenius}). Therefore, even moderate changes in $\Delta E$ lead to orders of magnitude variations in the lifetime at sufficiently low temperatures. Higher $\Delta E$ is commonly achieved by enhancing exchange frustration \cite{von2017enhanced,goerzen2023lifetime}, increasing DMI strength \cite{li2022strain}, or applying external magnetic fields \cite{romming2013writing}, all of which act to suppress skyrmion collapse by raising the activation energy. Another way to obtain higher lifetimes is to lower the pre-exponential factor $\Gamma_0$ \cite{Moritz_bimeron2025,shiwei_bimeron2025}, which describes entropic effects that can be strongly affected by the magnetic field \cite{Malottki2019}. Recent experimental measurements demonstrated that skyrmion lifetimes may be governed mainly by the entropic contribution to $\Gamma_0$ \cite{wild2017entropy}, leading to an unexpectedly short lifetime, even when the energy barrier is large. 
Long-lived magnetic solitons require simultaneously large energy barriers and suppressed entropic contributions to the pre-exponential factor.
However, a general mechanism capable of achieving this regime has remained elusive despite extensive efforts.

In this work, we uncover a previously unexplored stabilization mechanism in which engineering the geometry of the transition state enforces entropic suppression while preserving large energy barriers, yielding long and weakly temperature-dependent antiskyrmion lifetimes.
We show that aDMI provides a realistic route to realize spatially extended transition states in real materials.
To this end, we develop a first-principles method to compute aDMI for arbitrary NN pairs and apply it to investigate topological spin textures in oxidized Fe$_3$GeTe$_2$ (FGT-O), a material that has recently attracted intense experimental interest. Oxygen adsorption on a FGT monolayer breaks in-plane symmetry and induces a pronounced aDMI, which, together with a strongly reduced magnetocrystalline anisotropy energy (MAE), stabilizes nanoscale antiskyrmions with energy barriers exceeding 120 meV. More importantly, the aDMI gives rise to unconventional transition states at the SP, which present a larger spatial extension than those calculated for iDMI. These extended configurations preserve the Hessian zero modes inherent to the initial soliton, preserving the system's underlying translational symmetry. As a result, the entropic contribution to the lifetime prefactor is fully suppressed, yielding nearly temperature-independent lifetimes over a broad temperature range. When extrapolated to RT, the resulting lifetimes exceed those of state-of-the-art ultrathin transition-metal films by more than 5 orders of magnitude. 
Finally, we show that extended SPs are not unique to aDMI and identify the conditions for their emergence, establishing a general mechanism for soliton decay with temperature-independent prefactors.

\section{THEORY AND MODEL}

\subsection*{A. Toy model}
A magnetic atomic shell is defined as the set of magnetic lattice sites equidistant from a reference central site. For iDMI, all DMI vectors within a shell have identical magnitudes and follow a uniform rotational sense -- either clockwise or anticlockwise. In contrast, aDMI is characterized by site-dependent magnitudes and rotational senses within the same shell (Fig.~1 (a)). The topological nature of DMI-stabilized solitons is quantified by the topological charge:
\begin{equation}
Q=\frac{1}{4\pi} \int_{\mathbb{R}^2} \V{m} \cdot \left(\frac{\partial \V{m}}{\partial x} \times \frac{\partial \V{m}}{\partial y}\right) ~\mathrm{d}x\mathrm{d}y~,
\end{equation}
where $\textbf{m}$ is the unit vector parallel to the local magnetization, and $x$ and $y$ are Cartesian coordinates of the 2D lattice. Skyrmions and antiskyrmions carry opposite topological charges, $Q = -1$ and $Q = +1$, respectively. While iDMI naturally stabilizes skyrmions, antiskyrmions require aDMI. As illustrated in Fig.~2 (b), the iDMI energy density of a skyrmion is equally minimized in all angular directions, making it energetically favourable for iDMI. In contrast, an antiskyrmion possesses direction-dependent chiralities, resulting in an iDMI energy density that breaks in-plane rotational symmetry and can therefore be stabilized by certain types of aDMI.

\begin{figure}[t]
	\centering
	\includegraphics[width=1.0\linewidth]{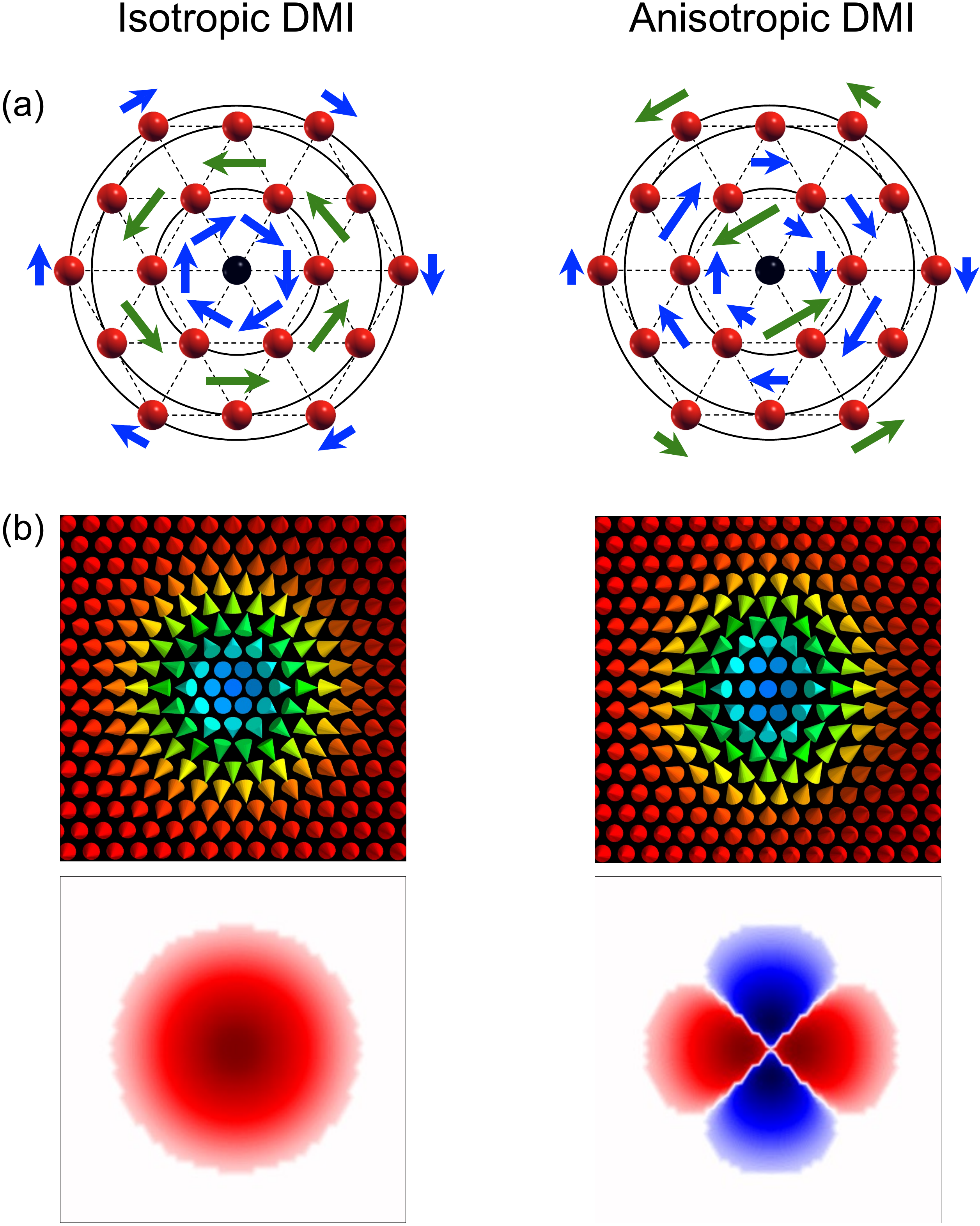}
	\caption{\label{Fig1} Isotropic versus anisotropic DMI. (a) Schematic representation of iDMI and aDMI vectors up to the $3^{\text{rd}}$ shell relative to the central reference site (black) on triangular magnetic site lattices. (b) Spin textures and spatial distributions of the iDMI energy density for a skyrmion ($Q=-1$) and an antiskyrmion ($Q=1$). The skyrmion exhibits a rotationally symmetric DMI energy density, whereas the antiskyrmion develops an anisotropic, multi-lobed pattern with alternating sign, showing direction-dependent chiralities.
    }
\end{figure}

To illustrate this interplay between the nature of DMI and the stabilized magnetic solitons, we employ a minimal model on a discrete two-dimensional (2D) lattice, restricted to a single shell. We consider a reference central spin with unit vector $\V{m}_{n} = \hat{\mathbf{z}}$ surrounded by $Z$ NNs. Their positions with respect to the central spin are specified by their polar coordinates ($\rho$, $\phi_{n'}$) 
and the spin directions are expressed in terms of the spherical angles $\Theta$ and $\Phi(\phi)$ as:
\begin{equation}
    \V{m}_{n'} = \left( \begin{array}{c}    
            \sin\Theta\cos\Phi(\phi_{n'})\\
            \sin\Theta\sin\Phi(\phi_{n'})\\
            \cos\Theta
            \end{array}  \right)~,\quad \Phi(\phi_{n'})=\nu\phi_{n'}+\gamma~.
\end{equation}
Here $\nu\in\mathbb{Z}$ is the vortex number, $\gamma\in[0,2\pi]$ is the helicity and $n'=1$,...,$Z$. The azimuthal angle function $\Theta=\Theta(\rho)$ is constructed to be constant for all spins $\mathbf{m}_{n'}$ in a given shell. The vortex number is related to the topological charge as $Q =  p\nu$ with polarity $p \in \{-1, 1\}$. Following Moriya's rules \cite{moriya1960anisotropic}, appropriate for the symmetry of the system that we are studying, the DMI vector  between center $n$ and the site $n'$ is perpendicular to the bond between the two sites, and lies in the shell plane:
\begin{equation}
    \mathbf{D}_{n n'} = D_{n n'}\left(\begin{array}{c}
            \cos (\phi_{n'} \pm  \pi/2)   \\
            \sin (\phi_{n'} \pm  \pi/2)  \\
            0
        \end{array} \right)~.
\end{equation}
The resulting total DMI energy reads:
\begin{equation}
\begin{aligned}
E_{\text{DMI}}
&= -\sum_{n n'} \mathbf{D}_{n n'} \cdot
   \left( \mathbf{m}_n \times \mathbf{m}_{n'} \right) \\
&= - \sum_{n n'}\,
   D_{n n'} \sin\Theta \sin\!\left[
      (1 - \nu)\phi_{n'} - \gamma \pm \pi/2 
   \right]~.
\end{aligned}
\end{equation}
The values of $D_{n n'}$ have been adjusted to avoid spurious double counting.
For iDMI, the contributions to $E_{\text{DMI}}$ must be identical from all sites within the shell. Also, this requirement enforces a constant DMI magnitude $D_{n n'}$ within the shell and stabilizes a radially symmetric soliton with uniform azimuthal angle  $\Theta$ and a single helicity $\gamma$. For $\nu=1$, corresponding to a skyrmion, the $\phi$ dependence of $E_{\text{DMI}}$ vanishes, rendering an identical $E_{\text{DMI}}$ contribution from all sites and thus explaining why skyrmions are stabilized by iDMI. In contrast, for $\nu=-1$, corresponding to an antiskyrmion, the angular dependence remains, preventing its stabilization by iDMI.

To illustrate this connection, we establish a general relation between the angular modulation of DMI and the vortex number. This establishes a unified language connecting lattice-level interactions and overall topology. For this, instead of treating the DMI vectors at each site independently, we assume that as the bond angle $\phi$ increases along the shell, the DMI vector rotates progressively, completing $f$ full rotations over $2\pi$. The additional phase shift $\phi_0$ accounts for an arbitrary global rotation offset.  At this stage, suppressing the Moriya's rule constraint, the DMI vector can be written as:
\begin{equation}
    \mathbf{D}_{n n'} = D_{n n'}\left(\begin{array}{c}
            \sin\Theta' \cos (f\phi_{n'} +  \phi_0)  \\
            \sin\Theta' \sin (f\phi_{n'} +  \phi_0)  \\
            \cos\Theta'
        \end{array} \right).
\end{equation}
We restrict ourselves to a radially symmetric soliton ($D_{n n'}$ does not explicitely depend on $n'$, but is rather constant within a given shell) with uniform helicity 
which already captures the essential physics. In the case of a single shell around a reference central spin, the total DMI energy then reduces to: 
\begin{equation}
\begin{aligned}
E_{\text{DMI}}
&= - D_{n n'} \sin\Theta' \sin\Theta
\sum_{n'=1}^{Z}
\sin\!\left[
(\nu-f)\phi_{n'} + \gamma - \phi_0
\right] \\
&= - D_{n n'} \sin\Theta' \sin\Theta
\sum_{n'=1}^{Z}
\cos\!\left[
(\nu-f)\frac{2\pi n'}{Z} + \xi
\right]~,
\end{aligned}
\label{eq:EDMI_first}
\end{equation}
where $\xi=\gamma-\phi_0-\frac{\pi}{2}$. Here, we take advantage of the fact that on a discrete Bravais lattice all lattice points within a shell can be addressed by $\phi_{n'}=2\pi n' /Z$. This is even true for, for example, the $4^{\text{th}}$ shell of the hexagonal lattice, which has 12 unequally spaced lattice points, but can be separated into two sets of equally spaced points.
For such an equally spaced set of lattice points, the summation in Eq.~(\ref{eq:EDMI_first}) resembles a geometric series and evaluates to:

\begin{equation}\label{eq:summation_dmi}
\sum_{n'=1}^{Z}
\cos\!\left[
(\nu-f)\frac{2\pi n'}{Z}+\xi
\right]
=
\begin{cases}
Z\cos\xi, & (\nu-f) \in Z\mathbb{Z} \\
0, & \text{otherwise}
\end{cases}~.
\end{equation}
Thus, within this idealized picture, a finite DMI contribution to a radially symmetric soliton arises only when the vortex number $\nu$ matches the number of DMI rotations $f$. Further, restoring Moriya’s rules corresponds to $\Theta'=\pi/2$ and projection of the DMI perpendicular to the bonds and in the shell plane. This simple single-shell model establishes a direct relation between the soliton topology and the angular modulation of the DMI. Each discrete shell carries a particular topological preference depending on DMI in that shell. In realistic systems, however, multiple shells contribute, each characterized by a different effective DMI rotation. This competition favors distinct soliton types across shells, rendering the identification of the most stable soliton a non-trivial problem that must ultimately be addressed through numerical simulations.

\subsection*{B. Atomistic spin Hamiltonian}

The energy of a magnetic configuration in real space is given by the extended Heisenberg model as:
\begin{equation}
\begin{split}
E & =-\sum_{nn'}J_{nn'}\V{m}_n \cdot \V{m}_{n'}-\sum_{nn'}\V{D}_{nn'} \cdot(\V{m}_n \times \V{m}_{n'}) \\
& -K \sum_n (m_n^z)^2 - \mu\sum_n (m_n^z B_z)~.
\end{split}
\label{Spin_model}
\end{equation}
These four terms correspond to Heisenberg exchange, DMI, MAE, and Zeeman energy under an external (out-of-plane) magnetic field $B_z$, respectively. These interactions are characterized by constants $J_{nn'}$, $\V{D}_{nn'}$, $K$, and $B_z$, while $\mu$ denotes the magnitude of the atomic magnetic moment. 
In most previous works, the DMI is assumed to be isotropic with respect to lattice directions. In the present work, however, this is not the case, necessitating a more advanced theoretical framework beyond the state-of-the-art approaches. We, therefore, adopt a formulation in reciprocal space that exploits the properties of cycloidal spin spirals.

For a magnetic configuration defined on a 2D Bravais lattice 
$L = \{ \mathbf{r}_n \mid \mathbf{r}_n = \sum_{i=1}^2 n_i \mathbf{a}_i, \, n_i \leq \sqrt{N}\ \}$
with $N$ discrete sites at positions 
$\mathbf{r}_n \in \mathbb{R}^2$ and lattice vectors $\mathbf{a}_i \in \mathbb{R}^2$, we employ a Fourier representation for magnetization characterized by wave vectors $\mathbf{q}_p \in \mathbb{R}^2$ $<$ and coefficients
$\mathbf{c}_p \in \mathbb{C}^3$:
\begin{equation}
\mathbf{m}(\mathbf{r}) = \sum_{p=1}^{N} \mathbf{c}_p e^{i \mathbf{q}_p \cdot \mathbf{r}}~.
\label{eq:fourier}
\end{equation}
Each coefficient can be written as $\mathbf{c}_p = \mathbf{R}_p + \mathrm{i} \mathbf{I}_p$ with
$\mathbf{R}_p$ and $\mathbf{I}_p$ representing its real and imaginary parts. The reciprocal lattice is defined as $L^* = \{\mathbf{q}_p \mid \mathbf{q}_p = \sum_{i=1}^2 p_i \mathbf{b}_i, \, p_i \leq \sqrt{N}\ \}$ 
A spin spiral is a special configuration constructed as a symmetric superposition of two waves with wave vectors $\mathbf{q}_p$ and $-\mathbf{q}_p$, and coefficients $\mathbf{c}_p$, $\mathbf{c}^{*}_p$, yielding the magnetization as:
\begin{equation}
\begin{split}
\mathbf{m}(\mathbf{r})
&= \mathbf{c}_p\, e^{i\mathbf{q}_p\cdot\mathbf{r}}
 + \mathbf{c}^{*}_p\, e^{-i\mathbf{q}_p\cdot\mathbf{r}} \\
&= 2\!\left[\mathbf{R}_p\cos(\mathbf{q}_p\cdot\mathbf{r})
 - \mathbf{I}_p\sin(\mathbf{q}_p\cdot\mathbf{r})\right]~,
\end{split}
\label{eq:spinspiral}
\end{equation}
with $|\mathbf{R}_p| = |\mathbf{I}_p| = 1/2$ and $\mathbf{R}_p \perp \mathbf{I}_p$. Distinct wave vectors $\mathbf{q}_p$ in the Brillouin zone (BZ) thus correspond to different spin spiral states. 

The Heisenberg exchange per site for a spin spiral with wave vector $\mathbf{q}$ can be expressed as:
\begin{equation}
E_{\text{exc}}(\mathbf{q}) = - \sum_{i=1}^{V} \sum_{j=1}^{Z_i} J_{i}^{j} \cos(\mathbf{q} \cdot \mathbf{r}_{j})~,
\label{Eq:Exchange_reciprocal}
\end{equation}
where $V$ denotes the number of shells included in the model, and $Z_i$ is the number of NN in the $i^{\text{th}}$ shell. $J_{i}^{j}$ coefficients take the same value for the magnetic sites located at $\mathbf{r}_{j}$ and $-\mathbf{r}_{j}$ for triangular lattices.
The corresponding DMI energy per site is written as:
\begin{equation}
E_{\text{DMI}}(\mathbf{q}) = - (\hat{\mathbf{q}} \times \hat{\mathbf{z}}) \cdot \sum_{i=1}^{V} \sum_{j=1}^{Z_i} D_{i}^{j} (\hat{\mathbf{r}}_{j} \times \hat{\mathbf{z}}) \sin(\mathbf{q} \cdot \mathbf{r}_{j})~,
\label{Eq:DMI_reciprocal}
\end{equation}
where the coefficients $D_{i}^{j}$ are the same for magnetic sites of triangular lattices located at $\mathbf{r}_{j}$ and $-\mathbf{r}_{j}$, due to symmetry considerations.
From Eqs.~\eqref{eq:spinspiral}, ~\eqref{Eq:Exchange_reciprocal}, and ~\eqref{Eq:DMI_reciprocal}, it follows that spin spirals are the general solutions for the Hamiltonian involving exchange and DMI terms, forming a natural basis for analyzing systems with arbitrary exchange and DMI.

\subsection*{C. First-principles calculations}

To determine the magnetic interaction constants in the system under study, we performed density functional theory (DFT) calculations. The exchange interactions require the energy dispersions of flat spin spirals calculated without SOC. As spin spirals break the translational symmetry of the lattice in the direction of $\mathbf{q}$, the Bloch theorem could only be applied by using the corresponding supercell, resulting in extremely time-consuming calculations. 
In the absence of SOC and for perfect lattices, all lattice sites are equivalent since they have identical local environments and magnetic moment magnitudes. This enables the use of the generalized Bloch theorem~\cite{kurz2004ab}, in which a lattice translation by a Bravais lattice vector $\mathbf{R}$ is supplemented by a spin rotation by an angle $\mathbf{q} \cdot \mathbf{R}$ about the spiral axis (normal to the spin-rotation plane). Equivalently, the spin spiral eigenstates can be written as:
\begin{align}
\psi_{\mathbf{k}n}(\mathbf{q},\mathbf{r})
&= \langle \psi_{\mathbf{k}n}(\mathbf{q}) | \mathbf{r} \rangle \nonumber \\
&= e^{i \mathbf{k}\cdot \mathbf{r}}
\begin{pmatrix}
e^{- \frac{i}{2}\mathbf{q}\cdot \mathbf{r}} & 0 \\
0 & e^{+ \frac{i}{2}\mathbf{q}\cdot \mathbf{r}}
\end{pmatrix}
\begin{pmatrix}
u^{\uparrow}_{\mathbf{k}n}(\mathbf{r}) \\
u^{\downarrow}_{\mathbf{k}n}(\mathbf{r})
\end{pmatrix}~.
\end{align}
Here, $\psi_{\mathbf{k}n}$ denotes the spin spiral eigenstates ($n$ being the band index) and $u_{\mathbf{k}n}$ are the Bloch functions. As a result, the spin spiral energy dispersion can be computed within the normal unit cell rather than the supercell.

The DMI energy was further computed using first-order perturbation theory \cite{heide2009describing}, based on self-consistently converged spin-spiral states with SOC as:
\begin{equation}
    E_{\text{DMI}}(\mathbf{q}) 
    = \sum_{\mathbf{k} n} 
    f_{\mathbf{k} n}(\mathbf{q}) \,
    \delta \epsilon_{\mathbf{k} n}(\mathbf{q})~,
    \label{eq:EDMI}
\end{equation}
where
\begin{equation}
    \delta \epsilon_{\mathbf{k}n}(\mathbf{q}) = \langle \psi_{\mathbf{k}n}(\mathbf{q}) | H_{\text{SOC}} | \psi_{\mathbf{k}n}(\mathbf{q}) \rangle ~,
    \label{eq:deltaE}
\end{equation}
$n$ denotes the band index and 
$f_{\mathbf{k} n}(\mathbf{q})$ is the occupation number of the scalar-relativistic spin spiral eigenstate $\psi_{\mathbf{k}n}$ computed without SOC. 

Finally, the MAE was determined self-consistently with SOC, by aligning the spin quantization axis along mutually perpendicular lattice directions. It is expressed as the total energy difference:
\begin{equation}
    E_{\text{MAE}}=  E^{\text{tot}}_\parallel - E^{\text{tot}}_\perp~,
\end{equation}
where $E^{\text{tot}}_\parallel$ and $E^{\text{tot}}_\perp$ are respectively the ground state energies computed for magnetic moments parallel and perpendicular to the 2D magnetic crystal.
DFT calculations were performed using the \textsc{fleur} code \cite{fleurv26} based on the full-potential linearized augmented plane wave (FLAPW) formalism. We used the local density approximation (LDA) with the parameterization proposed by Vosko, Wilk, and Nusair (VWN) \cite{vosko1980accurate} to include exchange-correlation effects. We did not include any on-site $U$ parameter as FGT-O is metallic and its electronic structure is not expected to be significantly affected by local correlations. Consistently, previous studies have shown that LDA accurately reproduces the experimental spin moments \cite{zhuang2016strong,deng2018gate}, while more recent work indicates that DFT+$U$ provides a less reliable description of the magnetic properties in the FGT family \cite{ghosh2023unraveling}. We used a cutoff parameter for the FLAPW basis functions of $k_{\text{max}}$ = 4.1  a.u.$^{-1}$, and we included basis functions up to spherical harmonics with $l_{\text{max}}$ = 8. The muffin tin radii used for Fe, Ge, and Te are 2.10 a.u., 2.10 a.u., and 2.63 a.u., respectively. In addition, we treated 3$s$, 3$p$, and 4$d$ states by local orbitals for Fe and Te, respectively. We have converged the total energy of flat spin spiral states using a $33 \times 33$ $\V{k}$-points mesh in order to obtain an accuracy of 0.001 meV. We used a denser mesh of $87 \times 87$ $\V{k}$-points for MAE calculations.

\subsection*{D. Parameterization of anisotropic DMI from first-principles}\label{parameterization}

The converged DFT energies of various spin spirals, sampled along the high-symmetry paths of the irreducible BZ, were fitted to corresponding terms of the spin model to extract the exchange and the DMI constants. In contrast to iDMI, which requires sampling along a single irreducible path within the first BZ, aDMI necessitates multiple paths of the same geometry along symmetry-inequivalent directions. In the present model of a triangular lattice, this entails three inequivalent ${\overline{\Gamma\mathrm{K}'\mathrm{M}'}}$ triangular paths for aDMI sampling. Note that ${\overline{\Gamma\mathrm{K}'\mathrm{M}'}}$ denotes sampling beyond the first BZ, which arises from the multilayer nature of FGT-O, as demonstrated in our previous work \cite{li2026stability}.

We now elaborate the fitting procedure for the case of aDMI. Here, Eq.~\eqref{Eq:DMI_reciprocal} is linear in DMI constants, and thus DMI contribution to the energy of a spin spiral can be expressed as:
\begin{equation}
E_{\text{DMI}}(\mathbf{q}) = {\mathbf{A}(\mathbf{q})} \mathbf{D}~,
\label{eq:linear_combination}
\end{equation}
where the row vector $\mathbf{A}(\mathbf{q})$ is given by:
\begin{equation}
\mathbf{A}(\mathbf{q}) = -2
\begin{bmatrix}
(\hat{\mathbf{q}} \times \hat{\mathbf{z}}) \cdot (\hat{\mathbf{r}}_1 \times \hat{\mathbf{z}}) \sin(\mathbf{q} \cdot \mathbf{r}_1) \\
(\hat{\mathbf{q}} \times \hat{\mathbf{z}}) \cdot (\hat{\mathbf{r}}_2 \times \hat{\mathbf{z}}) \sin(\mathbf{q} \cdot \mathbf{r}_2) \\
\vdots \\
(\hat{\mathbf{q}} \times \hat{\mathbf{z}}) \cdot (\hat{\mathbf{r}}_G \times \hat{\mathbf{z}}) \sin(\mathbf{q} \cdot \mathbf{r}_G)
\end{bmatrix}^{\mathsf{T}}~,
\end{equation}
and $\mathbf{D}$ is a column vector containing $G =\sum_{i=1}^{V} {Z_i}/2$ independent DMI magnitudes.
Spin spiral energies were computed for $s$ different $\mathbf{q}$-points distributed along the three non-equivalent ${\overline{\Gamma\mathrm{K}'\mathrm{M}'}}$ paths, 
resulting in a system of linear equations:
\begin{equation}
E_{\text{DMI}}(\mathbf{q}_t) = \mathbf{A}(\mathbf{q}_t) \mathbf{D}, \quad t = 1, 2, \dots, s.
\end{equation}
Since the number of sampled $\mathbf{q}$ vectors typically exceeds the number of unknown DMI constants, the system is overdetermined. To account for numerical noise in the DFT data and minor model mismatches, we solve the system using a least-squares approach. This procedure yields the DMI constants that best reproduce the DMI energies calculated from DFT, providing a statistically robust average fit across all sampled $\mathbf{q}$-points. We also applied a similar least-squares fitting method to obtain the exchange and the DMI constants for an isotropic lattice, using the corresponding high-symmetry path. DMI and exchange constants were extracted up to the $7^{\text{th}}$ and $8^{\text{th}}$ shells, respectively, ensuring quantitative convergence of the spin model. 

Beyond its immediate application to exchange and DMI, this comprehensive theoretical framework is fully general and can be straightforwardly extended to extract other complex magnetic interactions. As such, it offers a versatile pathway for systematically constructing accurate spin Hamiltonians directly from first-principles calculations, overcoming long-standing limitations of phenomenological modeling.

\subsection*{E. Implementation}

\begin{figure}[t]
	\centering
	\includegraphics[width=0.8\linewidth]{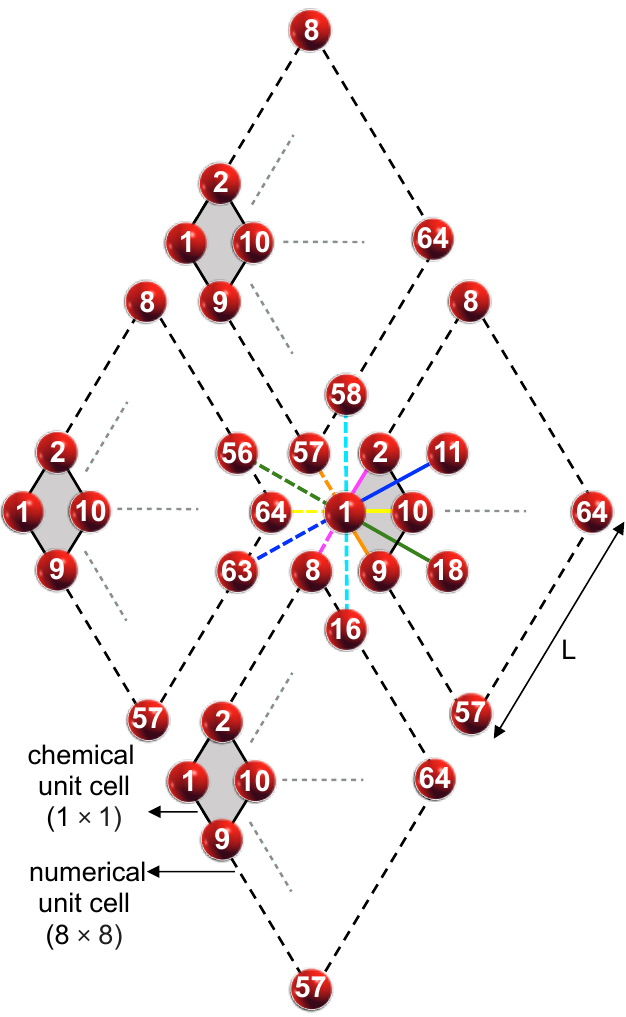}
	\caption{\label{spinakerunitcell} Unit cells in the \textsc{spinaker} code. The $1 \times 1$ chemical unit cell (grey), which is sufficient for describing iDMI, and the $8\times8$ numerical unit cell (dashed black lines), which ensures the computation of aDMI up to the $7^{\text{th}}$ shell without double assignment of interactions to identical pairs of lattice sites. This absence of contradictions is exemplarily illustrated at the example of the first two shells of interactions of the lattice site of species 1. It can be seen that all connections link the 1 to different species, which allows us to assign distinct DMI vectors to them. For reasons of lucidity, it is not shown that this uniqueness of assignments stays true for interactions up to the $7^{\text{th}}$ shell.
    }
\end{figure}

The atomistic spin simulations were performed with the \textsc{spinaker} code, using the spin Hamiltonian fully parametrized from first-principles calculations. 
In \textsc{Spinaker}, anisotropic bilinear interactions, such as DMI, are handled within a sufficiently numerical unit cell of size $L \times L$ containing $L^2$ lattice sites, usually much bigger than the chemical unit cell itself. In particular, we specify the set of anisotropic aDMI vectors $\{\mathbf{D}(\mathbf{r})\}$ which we obtained from \textit{ab initio} theory by the least-square method  described in the previous subsection (Sec. D). We explicitly take into account the antisymmetry $\mathbf{D}(-\mathbf{r}) = - \mathbf{D}(\mathbf{r})$. These interactions connect two lattice sites $\mathbf{r}_n$ and $\mathbf{r}_{n'}$, which can either be located within the same numerical unit cell or, because of periodic repetition, in adjacent cells. Because of this periodic construction, it can in principle occur that two lattice sites $\mathbf{r}_{n}$ and $\mathbf{r}_{n'}$ interact via two distinct DMI vectors $\mathbf{D}(\mathbf{r})\neq\mathbf{D}(\mathbf{r}')$, which fulfill $\mathbf{r}_{n'}= \mathbf{r}_n + \mathbf{r} = \mathbf{r}_n + \mathbf{r}' +\sum_{i=1}^2n_i\mathbf{A}_i$ where $\mathbf{A}_i=L\mathbf{a}_i$ is a unit vector of the numerical unit cell. Since this resembles an unphysical situation, \textsc{Spinaker} uses an approach from graph theory to find the minimal size $L\times L$ for the numerical unit cell, in which no such double assignment of interactions occurs and maps the set of DMI vectors $\{\mathbf{D}(\mathbf{r})\}$ to interactions between these $L^2$ lattice sites. Once this minimal size is found, a table of neighbors, which takes into account lattice sites within all periodically repeated unit cells, is created in a second step. This table is then iterated for the computation of energies, gradients and second derivatives of the Heisenberg model.

In our calculations, including exchange and DMI interactions up to the 7$^{\text{th}}$ coordination shell under aDMI, we determine that an $8\times8$ primitive supercell is the minimal size required to eliminate spurious self-interactions arising from periodic boundary conditions. This supercell, shown in Fig.~\ref{spinakerunitcell}, contains 64 lattice sites and 3518 distinct pairwise interactions.

\subsection*{F. Transition state theory}

To determine the metastable magnetic states corresponding to local minima of the energy landscape described by Eq.~(\ref{Spin_model}), we employed the velocity projection optimization (VPO) algorithm \cite{bessarab2015method}. The thermal stability of solitons was quantified by calculating the energy barriers protecting them against collapsing into the ferromagnetic (FM) ground state. The barrier was obtained using the geodesic nudged elastic band (GNEB) method \cite{bessarab2015method}, which allows for the determination of the MEP between distinct states. Along each MEP, a first-order SP was identified using the climbing-image procedure. The SP corresponds to the maximum energy $E_{\text{SP}}$ along the MEP, relative to the initial soliton state with energy $E_{\text{ini}}$. The energy barrier stabilizing a soliton against collapse is therefore given by: 
\begin{equation}
\Delta E = E_{\text{SP}} - E_{\text{ini}}~.
\end{equation}
Also, the image containing the Bloch-point-like (BP) texture was identified along MEP. Its position corresponds to the reaction coordinate between images for which the calculated topological charge $Q$ jumps from $\pm 1$ to $0$. 

To obtain a complete description of thermal stability, entropic and dynamical effects must also be taken into account. These contributions are quantified in the mean lifetime $\tau$ of a soliton, which within harmonic transition state theory (HTST) follows the Arrhenius law shown in Eq.~(\ref{tau_Arrhenius}).
Entropic effects are governed by the local curvature of the energy landscape in the vicinity of both the initial state and the SP, so not only
the initial soliton state and SP 
are considered, but also their thermal fluctuations. 
In HTST, a harmonic expansion of the energy around these extreme points is utilized:
\begin{equation}
E = E_{0} + \frac{1}{2} \sum_{n=1}^{2N} \lambda_{n} q_{n}^{2} + \mathcal{O}(q^{3})~,
\end{equation}
where $\lambda_{n}$ are eigenvalues of the Hessian at the initial state or the SP, and $q_{n}$ denote normal coordinates. 
The pre-exponential factor is given by \cite{bessarab2012harmonic}:
\begin{equation}\label{pf}
    \begin{split}
    {{\Gamma_0}} = {\frac{\Lambda}{2\pi} \left(2\pi k_{\text{B}}T\right)^{\frac{k_{\text{ini}}-k_{\text{SP}}}{2}} \frac{\prod_{n=2}^{1+k_{\text{SP}}} L_n^{\text{SP}}}{\prod_{n=1}^{k_{\text{ini}}}L_n^{\text{ini}}}
        \sqrt{\frac{\prod_{n=1+k_{\text{ini}}}^{2N}\lambda_n^{\text{ini}}}{\prod_{n=2+k_{\text{SP}}}^{2N}\lambda_n^{\text{SP}}}} }~,
    \end{split}
\end{equation}
where $\Lambda$ is the dynamical factor that contains the dynamic contributions from the
Landau-Lifshitz equation, $k_{\text{ini}}, k_{\text{SP}} \in \mathbb{N}$ are the number of zero modes, $\prod_{n=2}^{1+k_{\text{SP}}} L^{\text{SP}}_n$, ${\prod_{n=1}^{k_{\text{ini}}} L^{\text{ini}}_n}$ are the zero mode partition functions, and $\lambda_n^{\text{ini}}$ and $\lambda_n^{\text{SP}}$ are the eigenvalues of the Hessian of the respective states.

All atomistic spin simulations were performed using the \textsc{Spinaker} code. We constructed isolated skyrmions and antiskyrmions on field-polarized backgrounds, as well as their corresponding lattices, and fully relaxed all resulting spin textures. Note that, for the isotropic model of the FGT-O interactions, a lattice of $70 \times 70$ sites was sufficient for energy minimization. By contrast, in the anisotropic model, systematic convergence tests of the total energy with respect to lattice size demonstrate that a minimum lattice of $120 \times 120$ sites is required to ensure quantitative convergence for $B_z \geq 0.25$ T. At lower magnetic fields, significantly larger lattices of up to $500 \times 500$ sites are required to fully converge the calculations. Therefore, $\Delta E$ at lower $B_z$ was estimated via extrapolation. On the other hand, the lifetime calculations were therefore performed only for $B_z \geq 0.25$ T. In addition, the GNEB simulations were performed using a convergence torque of $10^{-8}$ eV and 25 intermediate images. We have verified that increasing the number of images (e.g., to 50 or 75) does not affect any of the results presented in this work.

\begin{figure*}[t]
	\centering
	\includegraphics[width=1.0\linewidth]{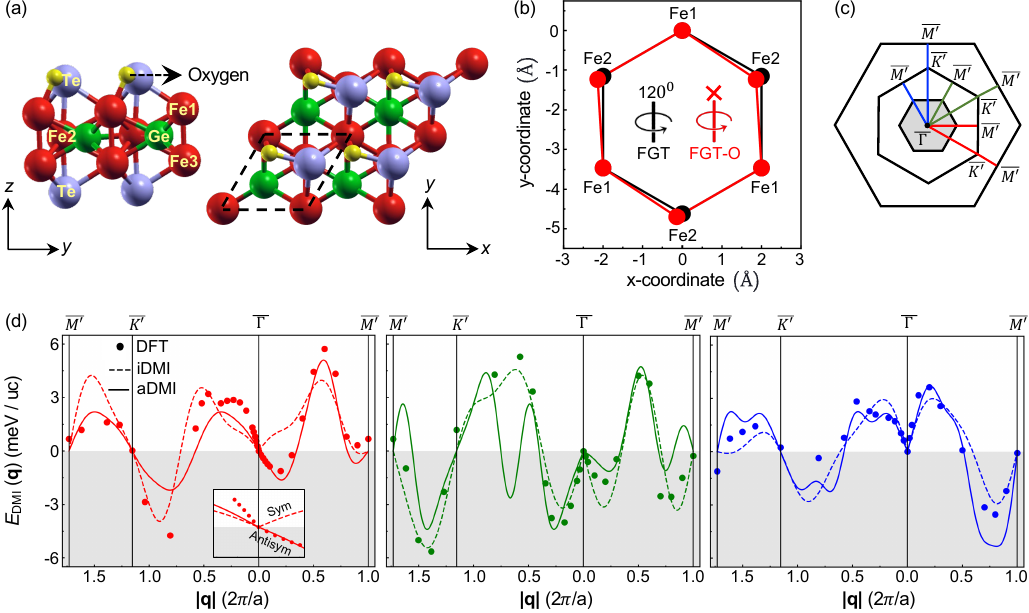}
	\caption{\label{Material_RMSD_SpinSpiral} (a) Side and top views of the atomic structure of monolayer FGT-O. The top, center, and bottom Fe atoms are labeled Fe1, Fe2, and Fe3, respectively. The black dashed lines indicate the 2D primitive unit cell. (b) Estimation of the degree of in-plane symmetry breaking for FGT-O. In this top view, Fe1 (on top of Fe3) and Fe2 atoms highlight an effective in-plane geometry. The black and red circles represent Fe atomic positions in FGT and FGT-O, respectively. The positions of Fe atoms in FGT are used as a reference for RMSD analysis. (c) Extended BZ with the first BZ highlighted in grey. Three high-symmetry paths along $\overline{\mathrm{M}'}$-$\overline{\mathrm{K}'}$-$\overline{\Gamma}$-$\overline{\mathrm{M}'}$are indicated: path 1 (red), path 2 (green), and path 3 (blue). (d) Energy dispersions of spin spirals per unit cell (uc) with SOC for FGT-O. Filled circles denote DFT total energies. Dotted lines represent fits to the DMI Hamiltonian within the isotropic (iso) model, while solid lines correspond to fits obtained under the anisotropic (aniso) model. For path 1 (red), a magnified view near the  $\overline{\Gamma}$ point highlights the antisymmetric (antisym) character of the DFT data and the anisotropic model fit, in contrast to the symmetric (sym) behaviour of the isotropic model fit (see inset). DMI interactions were included up to the $7^{\text{th}}$ shell. All energies are measured with respect to the FM state at the $\overline{\Gamma}$ point.}
\end{figure*}

\section{RESULTS}

\subsection*{A. Oxidized FGT monolayer}

We considered the FGT-O monolayer, an experimentally feasible platform, as a representative system for applying and validating our newly developed aDMI model. Freestanding monolayer FGT (space group P6$_3$/\textit{mmc}) is not a promising soliton material because of its inversion symmetry (no DMI) and strong MAE \cite{zhuang2016strong}. It has been shown that the surface of FGT rapidly oxidizes upon exposure to air~\cite{xie2024air}. Moreover, recent experiments demonstrated that exfoliation of FGT under non-vacuum conditions inevitably results in the formation of a surface oxide layer, which breaks inversion symmetry and restores DMI~\cite{gweon2021exchange,park2021neel}. Park \textit{et al.} \cite{park2021neel} experimentally demonstrated the formation of Néel-type magnetic skyrmions in FGT-O, which was explained using \textit{ab initio} calculations in terms of interfacial DMI. That analysis, however, was limited to the first NN and assumed a uniform rotational sense of the DMI. In reality, oxygen adsorption induces atomic displacements relative to pristine FGT, lowering the structural symmetry and providing a natural mechanism for aDMI with distinct rotational senses across different sites and shells. This anisotropy can substantially affect the morphology of solitons. Moreover, the experimental distinction between skyrmion and antiskyrmion textures remains challenging, further motivating our theory work. Below, we present the topological solitons stabilized under both iDMI and aDMI, going beyond the NN approximation.

We used the \textsc{Quantum Espresso} code \cite{fleurv26} for atomic relaxations of free-standing FGT and FGT-O monolayers. We considered van der Waals interactions using semi-empirical dispersion corrections (DFT-D2) formulated by Grimme \cite{grimme2010consistent}. We fully relaxed the systems until atomic forces became lower than 10$^{-6}$ Ry/Bohr. We identified five energetically stable FGT-O configurations corresponding to distinct O adsorption sites, including two bridge and three on top geometries.
In the following, we focus on the lowest energy configuration among these, which is consistent with the structure reported in Ref.~\cite{park2021neel} [Fig.~\ref{Material_RMSD_SpinSpiral} (a)]. The lattice constants calculated for the monolayers are about 4.00~\AA, in good agreement with experimentally reported values of about 3.99~$\sim~$4.03~\AA~\cite{chen2013magnetic}. Both pristine and oxidized monolayers retain a metallic character, as confirmed by the projected density of states (PDOS) analysis. Notably, direct bonding between Fe atoms and oxygen leads to pronounced modifications of the electronic structure near the Fermi energy ($E_{\text{F}}$), indicating that oxygen adsorption significantly tunes the magnetic interactions in FGT (see Fig.~\ref{pdos} in Appendix A).

\subsection*{B. Microscopic origin of anisotropic DMI}

The emergence of aDMI in FGT-O is rooted in the oxygen-induced breaking of in-plane rotational symmetry (see Fig.~\ref{Material_RMSD_SpinSpiral}(a-b)). While pristine FGT preserves 3-fold rotational symmetry in the basal plane, which enforces an isotropic form of the DMI, oxygen adsorption lifts this constraint by inducing in-plane lattice distortions that lower the rotational symmetry. As a result, $\mathbf{D}_{nn'}$ depends on the bond orientation and does not transform isotropically under in-plane rotations. To quantify the degree of in-plane symmetry breaking, we evaluate the root mean square deviation (RMSD) of the in-plane Fe atomic positions relative to pristine FGT, defined as:
\begin{equation}
\mathrm{RMSD} =
\sqrt{
\frac{1}{N}
\sum_{i=1}^{N}
\left|
\mathbf{r}_i - \mathbf{r}_i^{\mathrm{FGT}}
\right|^2
}~.
\end{equation}
where $N$ is the total number of Fe atoms under consideration, $\mathbf{r}_i$ denotes the position of the $i^{\text{th}}$ atom in the oxidized structure, and $\mathbf{r}_i^{\text{FGT}}$ is the corresponding atomic position in pristine FGT. 
The resulting RMSD of 0.15~\AA~indicates a substantial in-plane lattice distortion induced by oxygen adsorption. Such distortions render the otherwise crystallographically equivalent directions inequivalent, thereby removing the rotational symmetry constraints. Within the Fert-Lévy model \cite{Fert1980}, the DMI between Fe moments is mediated by SOC associated with the heavy Te atoms. Oxygen-induced in-plane displacements of Fe atoms modify the relative distances and bonding geometries between a given Fe atom and its surrounding Te and Ge ligands in a direction-dependent manner. As a result, the SOC-driven hopping processes become anisotropic, leading to a crystallographic direction dependence of $\mathbf{D}_{nn'}$. In the following, we analyze the magnetic interactions and the resulting topological spin textures, with a particular focus on the aDMI.

\subsection*{C. Spin spiral dispersions}

FGT has a complex structure in which three non-equivalent Fe atoms form an intrinsic multilayer configuration. Calculating its intralayer and interlayer interactions is, in principle, feasible, as we demonstrated in our previous work \cite{li2023tuning}, but it requires substantial computational effort. On the other hand, we have applied an effective collective 2D model in which the three Fe atoms are treated as a single magnetic unit \cite{li2022strain,Dongzhe2024}. This approach predicts mesoscopic properties such as topological spin textures and the Curie temperature, in good agreement with the literature on the FGT family of materials. Therefore, in this work, we apply this collective 2D model with a particular focus on aDMI.
 
We focus on spin spiral vectors $\mathbf{q}$ along the three high-symmetry directions ${\overline{\Gamma\mathrm{M}'}}$ and ${\overline{\Gamma\mathrm{K}'\mathrm{M}'}}$ of the BZ [Fig.~\ref{Material_RMSD_SpinSpiral}(c)]. Note that sampling $\mathbf{q}$ vectors beyond the first BZ is required due to the multilayer nature of FGT, as discussed in detail in Ref.~\cite{li2026stability}.
The high-symmetry points represent special magnetic states: $\overline{\Gamma}$ corresponds to the FM state, $\overline{\text{K}^{\prime}}$ to the N\'eel-state with an angle of 120$^{\circ}$ between adjacent spins, and $\overline{\text{M}^{\prime}}$ to the row-wise antiferromagnetic (AFM) state. 

We fit our DFT total energies to two types of models, namely an isotropic model and an anisotropic model. Within the isotropic approximation, the exchange Hamiltonian was fitted independently along each of the three high-symmetry paths: path 1 (red), path 2 (green), and path 3 (blue). 
We find that the exchange energies are nearly identical along the three paths in the vicinity of $\overline{\Gamma}$, indicating that the isotropic model already provides a good description of the exchange interaction (see Fig.~\ref{Exchange_fit} in Appendix B). The near coincidence of the three dispersions around $\overline{\Gamma}~(\mathbf{q}=0$) demonstrates that the long-wavelength exchange stiffness is effectively isotropic and therefore captures the dominant physics governing skyrmion stability. The deviations that develop at larger $\mathbf{q}$ reflect a residual anisotropy beyond the continuum limit. While this anisotropy does not qualitatively affect the existence of solitonic states, it renormalizes short-length-scale properties, including the core structure, equilibrium size, and collapse barrier, and thus becomes relevant primarily for compact skyrmions or strongly localized noncollinear textures. Note that, as we have confirmed, using exchange parameters extracted along any of the three paths does not alter the central conclusions of this work.

For the anisotropic model, the exchange parameters were instead obtained by averaging over the three isotropic fits. In the second step, the SOC contribution to the energy dispersion of spin spirals was fitted to the DMI Hamiltonian to extract the DMI parameters for arbitrary NN [Fig.~\ref{Material_RMSD_SpinSpiral}(d)]. In contrast to the exchange interaction, the DMI contribution exhibits a pronounced dependence on the selected high-symmetry path, directly reflecting the strong in-plane symmetry breaking induced by oxygen adsorption in FGT. The isotropic model fits the DFT data well along path 2 (green) and path 3 (blue). However, along path 1 (red), it fails completely near the $\overline{\Gamma}$ point, as it cannot capture the antisymmetric character of the DFT dispersion [see inset of Fig.~\ref{Material_RMSD_SpinSpiral}(d)]. This breakdown can be traced to the constraint imposed by iDMI, which enforces identical DMI behavior across all three high-symmetry paths and thus lacks the flexibility required to capture directional variations. In contrast, aDMI, obtained from a global fit to the DFT data along all three high-symmetry paths, faithfully reproduces the spin spiral dispersions over the entire BZ. This highlights the central role of the anisotropic model for FGT-O, which goes beyond the conventional iDMI framework. All exchange and DMI interaction parameters for both the isotropic and the anisotropic models and aDMI energy map are displayed in Fig.~\ref{aDMI_analysis} in Appendix B.

In addition, SOC gives rise to a finite MAE. We find that oxygen adsorption dramatically alters the MAE of FGT. For FGT-O, the strong out-of-plane MAE of -3.01 meV/uc in pristine FGT decreases significantly to -1.30 meV/uc, resulting in a moderate out-of-plane MAE.

The emergence of strong DMI together with a substantially reduced MAE provides a promising platform for the stabilization of topological spin textures.

\subsection*{D. Antiskyrmions driven by anisotropic DMI}

\begin{figure}[t]
	\centering
	\includegraphics[width=1.0\linewidth]{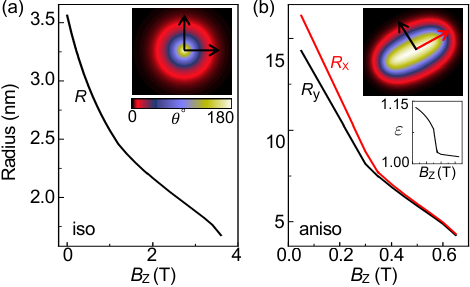}
	\caption{\label{antisky_size} Soliton size and deformation. (a) Skyrmion radius as a function of magnetic field $B_z$ in the isotropic model.
(b) Same as (a), but for the anisotropic model. Note that we used MAE = $K/10$ for the anisotropic model. The antiskyrmion becomes elliptically distorted due to aDMI. The radii are therefore evaluated along the principal axes of the ellipse and shown in two colors. Insets illustrate the $\theta$-profiles of solitons, highlighting the circular (isotropic) and elliptical (anisotropic) symmetry of the solitons. For antiskyrmions, we additionally plot the degree of elliptic deformation, defined by $\varepsilon = {R_x}/{R_y}$ as a function of $B_z$.}
\end{figure}

We first used the isotropic model approximation, extracting interaction parameters by fitting the DFT data collected along path 3 only [dotted blue curve shown in Fig.~\ref{Exchange_fit} of Appendix B and Fig.~\ref{Material_RMSD_SpinSpiral}(d)]. We have verified that qualitative conclusions on the physical properties further computed with these parameters are insensitive to this choice of a high-symmetry path, although the chirality of the resulting solitons is set by the sign of the DMI that can depend on the selected direction. 
Interestingly, the isotropic model stabilizes highly compact (sub-10 nm) Néel-type skyrmions, with radii below 4 nm already at zero magnetic field, as shown in Fig.~\ref{antisky_size} (a). Note that the soliton radius is evaluated using the standard definition introduced by Bogdanov \textit{et al.}~\cite{bocdanov1994properties, arya2025new}.

\begin{figure*}[t]
	\centering
	\includegraphics[width=1.0\linewidth]{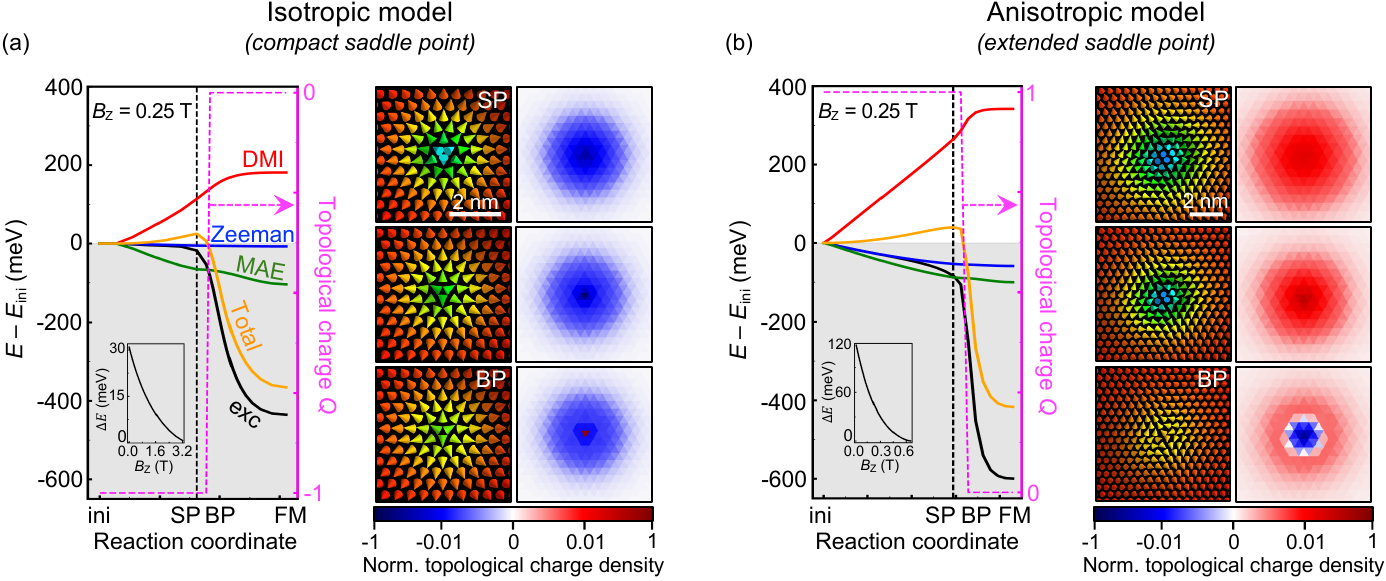}
\caption{MEP and SP character in the isotropic and the anisotropic models.
(a) Left: MEP for the collapse of a skyrmion initial (ini) state into the FM state at $B_z = 0.25\,\mathrm{T}$ within the isotropic model. The energy is decomposed into exchange (black), DMI (red), MAE (green), and Zeeman (blue) contributions. The SP is indicated by the vertical black dashed line. The topological charge $Q$ (right axis, magenta) identifies the BP. Center: spin configurations and right: corresponding normalized topological charge densities around the SP, the BP and their intermediate state. SP reveals a compact, radially symmetric instability characteristic of localized collapse. (b) Same as (a), but for the anisotropic model. In contrast to the isotropic case, the SP is spatially extended and accompanied by a broadly distributed topological charge density, leading to a collective, extended instability induced by aDMI. This qualitative change in the nature of the transition state underlies the distinct decay mechanism in the anisotropic system. Note that the 2 nm scale bar represents different physical lengths in (a) and (b). Insets show the field dependence of the energy barrier $\Delta E$.}
\label{gneb_iso_aniso}
\end{figure*}

A qualitatively different behavior emerges when the full anisotropic model is considered. In this case, the energetically preferred topological excitation is an antiskyrmion. While antiskyrmion latticees can also be stabilized, they are slightly higher in energy (by $\sim$0.1 meV per atom) than isolated antiskyrmions. Owing to the relatively strong MAE in FGT-O, antiskyrmions become stable for magnetic anisotropy values reduced below $K/5$. Upon decreasing the magnetic anisotropy, the antiskyrmions develop a pronounced elliptical distortion, which necessitates a direction-dependent evaluation of their spatial extent. In the following, we fix MAE to $K/10$, a value that is experimentally tunable through doping or temperature control ~\cite{tan2018hard, Wang2020modifications, park2019controlling}. Importantly, our main conclusions remain unchanged for other representative values, such as $K/5$ or $K/20$. Without applying out-of-plane magnetic ﬁeld ($B_{z}=0$ T), a stripe-like phase nearly degenerate with the FM state is observed, as discussed in Fig.~\ref{rad_barr_minima} in Appendix C. Upon increasing $B_{z}$, isolated elliptical antiskyrmions nucleate in the FM background. Remarkably, the antiskyrmion phase persists over a wide magnetic-field range, extending up to 0.65~T, beyond which the magnetic field-polarized FM phase is restored. The antiskyrmions are nanoscale throughout this range, with their characteristic size continuously decreasing from 16.2~nm to 4.2~nm as the magnetic field is increased [see Fig.~\ref{antisky_size}(b)].

We also depict $\theta$-profiles in the insets of Fig.~\ref{antisky_size} for both the isotropic and the anisotropic solitons. These maps represent the angle $\theta$ between each local spin $\mathbf{m}(\mathbf{r})$ and the reference FM background spin with orientation $\hat{\mathbf{e}}_{\text{FM}}$, defined as $\theta(\mathbf{r}) = \arccos \left( \mathbf{m}(\mathbf{r}) \cdot \hat{\mathbf{e}}_{\mathrm{FM}} \right)$. The black region corresponds to the FM background where the out-of-plane magnetization component satisfies $\theta=0$, while the white region indicates spins antiparallel to the FM background magnetization, corresponding to the soliton core. In the isotropic model, the skyrmion radius $R$ is uniquely defined, and the $\theta$-profile is radially symmetric due to rotational symmetry enforced by iDMI. As the magnetic field $B_z$ increases, $R$ decreases, reflecting the uniform compression of the skyrmion by the Zeeman interaction. In contrast, in the anisotropic model, the antiskyrmion lacks rotational symmetry, assuming an elliptical shape. To quantify its size, we determine two characteristic radii, $R_x$ and $R_y$, along the principal axes of the ellipse. These axes are identified from the real-space spin texture as the directions of maximal and minimal spatial extent of the antiskyrmion core. The ellipticity of the antiskyrmion is further characterized by the ratio, $\varepsilon = {R_x}/{R_y}$, which equals unity for a circular texture and deviates from unity in the presence of anisotropy. The field dependence of $R_x$ and $R_y$ indicates that aDMI not only stabilizes antiskyrmions but also controls their geometric anisotropy. At low magnetic fields, the antiskyrmion shape directly reflects the effect of aDMI. With increasing $B_z$, the Zeeman energy progressively dominates over the aDMI energy, shrinking the antiskyrmion and suppressing directional differences, which drives a crossover from an elliptical to an effectively isotropic texture [see the inset in Fig.~\ref{antisky_size}(b)].

\begin{figure*}[t]
	\centering
	\includegraphics[width=1.0\linewidth]{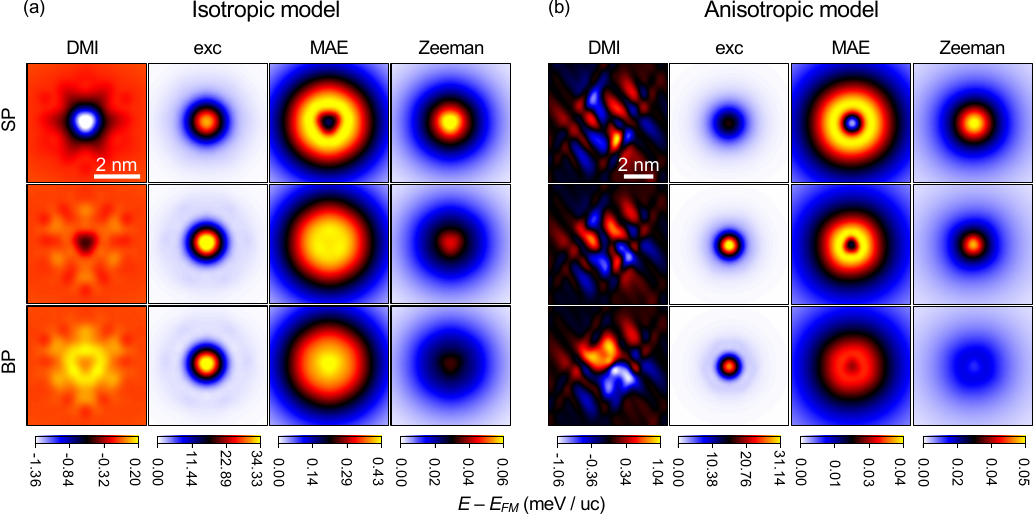}
\caption{Energy density maps for the isotropic and the anisotropic models. (a) Interaction-resolved energy density maps for selected configurations along the MEP for the isotropic model. The top, center, and bottom rows correspond to the SP, an intermediate state, and the BP configurations, respectively.
(b) Similar analysis for the anisotropic model during antiskyrmion collapse. All energies are given with respect to the FM state. The SP remains spatially localized in the isotropic model, whereas aDMI gives rise to a pronounced spatial delocalization of the SP in the anisotropic model. This clearly demonstrates that the extended character of the SP is a direct consequence of aDMI. The 2 nm scale bar represents different physical lengths in (a) and (b).}
\label{dmi_energy_density}
\end{figure*}

Then, the MEPs for the processes of the soliton collapse into the FM state were calculated using the GNEB method for both the isotropic and the anisotropic models. Results are shown in Fig. \ref{gneb_iso_aniso}. The FM state constitutes the ground state, both in the isotropic and anisotropic models. For both models, the DMI (red curve in the MEP energy decomposition plots) is the main ingredient that favors the formation of solitons. The GNEB calculations provide direct insight into the microscopic mechanisms governing magnetic transitions. 
Analysis of spin configurations and topological charge densities in the vicinity of the SP and the BP reveals a radial collapse mechanism for skyrmions, characterized by a symmetric contraction of the soliton prior to its annihilation into the FM state~\cite{muckel2021experimental} (see Fig.~\ref{gneb_iso_aniso}(a), right, for the corresponding spin textures and topological charge densities). By contrast, the antiskyrmion (Fig.~\ref{gneb_iso_aniso}(b), right) does not undergo radial shrinking but instead collapses anisotropically along its principal axes. In addition to this, a qualitative distinction emerges in the nature of the SPs. In the isotropic model, the SP exhibits the expected compact character -- the skyrmion is already strongly contracted, its core is highly compressed, bringing the spin texture to the brink of a localized instability. In contrast, within the anisotropic model, the SP remains spatially extended, largely preserving the characteristic structure of the initial nanoscale solitons reported in most systems . We emphasize that extended SPs were reported in quantum field theory a few decades ago \cite{Callan1977}, where decay proceeds via spatially extended bounce solutions, but have not been reported in the context of magnetic solitons, whose collapse is typically mediated by compact, localized transition states. 

The origin of the extended SP lies in the anisotropic helicity structure possessed by the antiskyrmion. Owing to this, energy costs are not uniform across all directions during a collapse. Instead, chiral rotations develop preferentially along the axes dictated by aDMI. To minimize the DMI energy penalties caused by conflicting chiral distortions, the instability does not remain localized; it instead spreads across an extended spatial region. This behavior can be understood within a micromagnetic description by decomposing the DMI energy density into directional gradient contributions:
\begin{equation}
\mathcal E_{\mathrm{DMI}}
=
D_x\, \mathcal F_x(\mathbf m,\partial_x \mathbf m)
+
D_y\, \mathcal F_y(\mathbf m,\partial_y \mathbf m)~.
\label{eq:dmi_aniso}
\end{equation}
Here $D_x $ and $D_y$ denote the DMI strengths along the $x$ and the $y$ axes, respectively, $\partial_x \mathbf m$ and $\partial_y \mathbf m$ describe spatial variations of the magnetization field $\mathbf{m}$, while $\mathcal F_x$ and $\mathcal F_y$ represent the corresponding chiral gradient contributions associated with modulations along the two in-plane directions. 
For the isotropic case, $D_x = D_y$, and the corresponding DMI Hamiltonian remains invariant under in-plane rotations generated by the operator $\hat{R}_z$, which rotates the spatial coordinates and the magnetization field about the out-of-plane $z$-axis. A compact, radial SP necessitates comparable gradient contributions along both axes. Within the core region of such a localized state:
\begin{equation}
\mathcal F_x(\mathbf m,\partial_x \mathbf m)
=
\mathcal F_y(\mathbf m,\partial_y \mathbf m)~,
\end{equation}
reflecting the isotropic distribution of spatial variations. As a result, $\mathcal E_{\mathrm{DMI}}$ varies without a favorable direction, and the energy cost of shrinking remains uniform in all directions. In contrast, for the anisotropic case ($D_x \neq D_y$), the DMI Hamiltonian no longer commutes with $\hat{R}_z$, and hence the symmetry is reduced. If, for instance, $D_x > D_y$ (with $D_x < 0$), the system gets a massive discount on its energy for every bit of twisting it does along the $x$-axis. The system stretches out along $x$ to gather as much DMI energy as possible over a wider area. Twisting along the $y$-axis is expensive in comparison, and hence it will resist twisting along this direction. The DMI energy, therefore, weighs directional chiral twists unequally. At SP, it refuses to shrink easily because that would lose too much stable energy. The system instead lowers its energy by relaxing the constraint of radial localization and spreading the instability over a larger spatial region. This extended deformation enables an anisotropic redistribution of gradients such that:
\begin{equation}
\mathcal F_x(\mathbf m,\partial_x \mathbf m)
>
\mathcal F_y(\mathbf m,\partial_y \mathbf m)~.
\end{equation}
The situation is reversed when $D_x < D_y$. In this sense, the emergence of an extended SP is a direct consequence of the directional selectivity imposed by aDMI, which is incompatible with a compact, radially symmetric instability. This distinction is directly manifested in the interaction-resolved energy density profiles along the MEP in the vicinity of SP for FGT-O, shown in Fig.~\ref{dmi_energy_density}. For the isotropic model, all interaction contributions remain spatially localized around the SP [Fig.~\ref{dmi_energy_density} (a)]. In contrast, in the case of the anisotropic model, aDMI enforces a strongly delocalized SP, with energy density extending over a large region of the system [Fig.~\ref{dmi_energy_density}(b)]. 

Note that the spatial extent of the SP also depends on the initial size of the antiskyrmion stabilized by aDMI and the relative slopes of the interaction curves along MEP during collapse. If the metastable antiskyrmion is already small, aDMI still imposes anisotropic deformation during the collapse process, maintaining texture elongation along the energetically favored axis. In this case, however, the resulting SP is less spatially extended than in the scenario discussed above. Nevertheless, it remains significantly more delocalized than the corresponding SP obtained in the isotropic model with interaction parameters of similar magnitude. 
Furthermore, the variation of magnetic interactions along the MEP must be relatively balanced. If the slope of one non-chiral interaction curve changes drastically, it may overpower the others, potentially suppressing the anisotropic features. These considerations indicate that the emergence of a sufficiently extended SP requires not only strong aDMI, but also a large enough initial antiskyrmion size and stable interaction ratios throughout the collapse. Even if we have a potential system that satisfies the prerequisites to possess aDMI, but the metastable soliton is small, we can still harness the aDMI benefits of extended textures. By using experimental techniques to tune the ratio of DMI to other interactions, we can effectively manipulate the textures to maintain delocalization.

Further, the calculated annihilation barrier for FGT-O differs by more than a factor of 4 between the two models. In the isotropic case, the skyrmion collapse barrier is $\sim$30~meV at zero field, whereas in the anisotropic case, the antiskyrmion barrier reaches $\sim$120~meV at finite field (see insets in Fig.~\ref{gneb_iso_aniso}). This substantially larger $\Delta E$ for antiskyrmions compared to skyrmions originates from their larger texture size. Since the collapse entails a collective rearrangement of spins over the entire soliton area, a larger soliton incurs a higher total energy cost, resulting in an increased $\Delta E$. Such a large antiskyrmion barrier is comparable to those in state-of-the-art ultra-thin films considered as prototype systems for hosting nanoscale solitons \cite{von2017enhanced,Haldar2018}.

Taken together, these findings reveal that an analysis constrained to iDMI fails in capturing the essential physics in systems with reduced in-plane symmetry. In FGT-O, the bond dependence of DMI plays a crucial role in governing both the nature and the stability of topological solitons. A description based on aDMI is therefore indispensable for explaining the emergence of antiskyrmions and for a faithful representation of the underlying energy landscape. More generally, we find that aDMI is not the only route to extended transition states. In the following section, we identify the general conditions under which such saddle points emerge.

\subsection*{E. Mechanisms for the emergence of extended saddle points}

To clarify the origin of the extended SP, we first analyze the conventional central collapse of a topological soliton. Along the MEP, the system typically evolves from an extended initial through increasingly compact configurations before collapsing into the FM state (except when the initial state is already compact). In this context,, the extended configuration is an intrinsic stage along the collapse trajectory, with the SP corresponding to the energy maximum along the path. Consequently, the emergence of an extended SP requires that the energy of the extended configurations exceeds that of compact ones, effectively shifting the barrier toward the delocalized regime.

In the iDMI case [Fig.~\ref{gneb_iso_aniso} (a)], the energy landscape is dominated by the monotonic increase of the DMI contribution on the extended side of the MEP (left of the SP). In this region, non-chiral terms, including exchange, MAE, and Zeeman energy, offer only a marginal contribution, resulting in a compact SP. When aDMI is introduced [Fig.~\ref{gneb_iso_aniso}(b)], the energy balance is reshaped. The exchange interaction provides a significant negative contribution much earlier in the collapse process. Crucially, on the compact side of the SP, the reduction in relative energy $(E-E_{\text{ini}})$ from non-chiral terms outpaces the increase in DMI energy along the MEP.
This preferential lowering of the compact region relative to the extended regime effectively shifts the energy maximum, resulting in an extended SP.

\begin{figure}[t]
	\centering
	\includegraphics[width=0.99\linewidth]{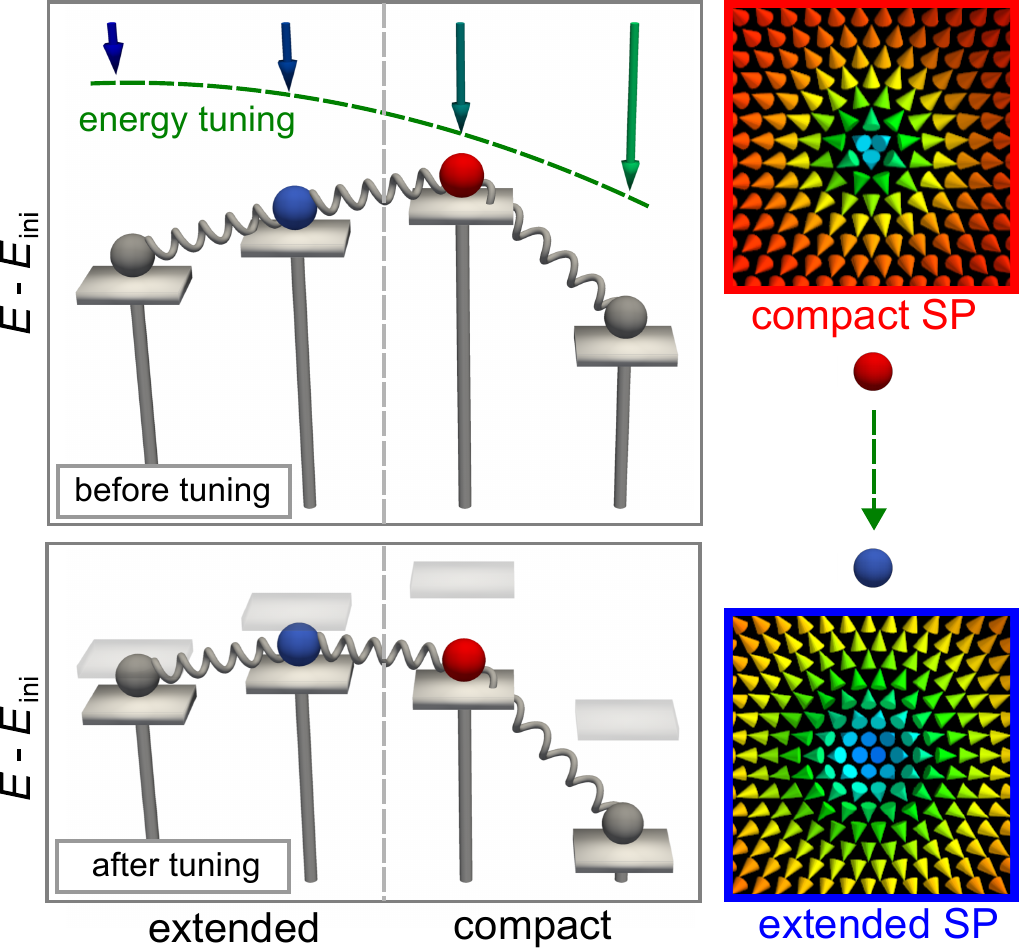}
	\caption{\label{SP_transition} Schematic illustration of an alternative route to obtain an extended SP via energy tuning.
Before tuning, the energy maximum along the MEP is located in the compact configuration region, leading to a compact SP (red sphere).
Due to the stronger spin variations in this region, the energy landscape around this SP is relatively steep.
The arrows in the upper panel indicate the energy changes of individual images relative to the initial state induced by energy tuning. For example, increasing the magnetic field can provide a larger energy compensation for compact configurations compared to extended ones.
The transparent platforms in the lower panel mark the energy profile before tuning. After energy tuning, the energy maximum point shifts toward the extended configuration region, resulting in the emergence of an extended SP (blue sphere) and a flatter energy landscape in its vicinity.
    }
\end{figure}

Besides, a similar SP shift can be achieved by directly tuning the energy landscape, as illustrated in Fig.~\ref{SP_transition}. 
In the upper panel, downward arrows denote the energy variations of individual images along the MEP induced by enhancing the non-chiral interactions or reducing the DMI strength—both of which provide greater energy compensation for compact configurations. This selective lowering of the compact region's energy, relative to the comparatively less affected extended region, modifies the overall energy profile. Consequently, as shown by the contrast with the initial state (transparent platforms), the maximum along the MEP shifts from the compact to the extended configuration, giving rise to an extended SP.

However, such direct energy tuning typically leads to a reduction of the skyrmion size, which in turn lowers the energy barrier and reduces the overall stability. In this context, aDMI provides a more advantageous route, as it enables a rebalancing of energy contributions while simultaneously allowing the skyrmion size to increase. 
These results highlight the unique role of aDMI. In anisotropic systems, extended SPs are symmetry-enforced by the presence of aDMI, where its isotropic model counterpart would remain
compact. On the other hand, in isotropic systems, they arise exclusively from interaction-specific energy tuning. Importantly, these two mechanisms are not mutually exclusive for aDMI. Even in the presence of aDMI, the energy landscape can be further tuned, for example, by varying the magnetic field.

Beyond extended SP induced by aDMI in FGT-O, we investigated other isotropic physical systems to contextualize these findings. Specifically, we examined the Janus monolayer MnSeTe \cite{arya2025new}, which hosts extended magnetic textures at zero field at both the
initial metastable skyrmion and the SP states.
In MnSeTe, these textures are large due to parameters only
in the limit of zero external field. However, applying a finite field rapidly drives the system into a regime where the SP becomes spatially compact. On the other hand, aDMI in FGT-O
forces the instability to remain spatially delocalized regardless
of the field strength. This suggests that while isotropic solitons are inherently fragile – undergoing transition to a compact SP upon parameter tuning, anisotropic solitons exhibit a
structural robustness. In these systems, the extended nature
of the SP is not a consequence of parameter proximity, but a
direct manifestation of the underlying aDMI symmetry.

In the following section, we further examine the thermal and entropic stability of antiskyrmions in the presence of extended transition states, which play a crucial role in their potential use in device applications.
 
\subsection*{F. Lifetime of antiskyrmions}

\begin{figure}[t]
	\centering
	\includegraphics[width=0.93\linewidth]{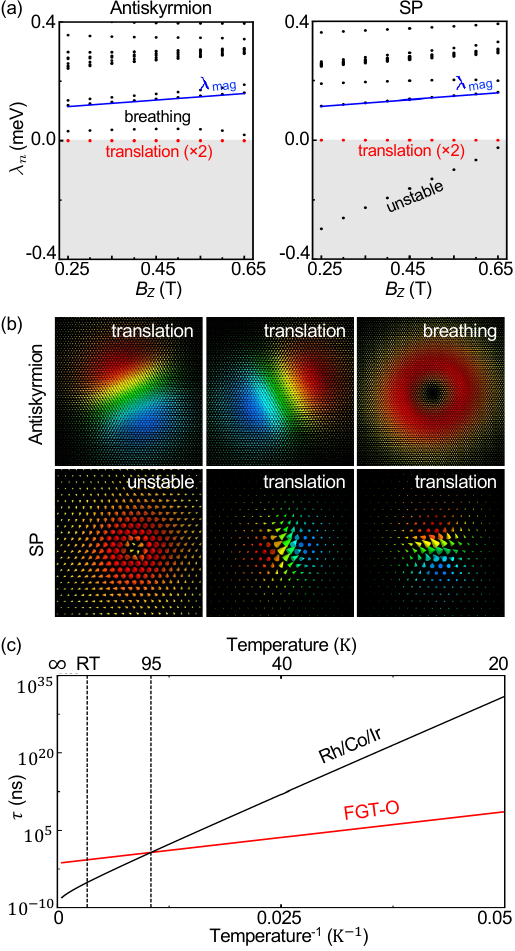}
	\caption{\label{HTST} (a) Hessian eigenvalue spectra of the antiskyrmion and the SP states over magnetic fields $B_z$. Lifetimes were evaluated for $B_z > 0.25$ T as the substantially larger lattice sizes required at lower fields render the calculations too computationally demanding. Only low-energy modes are shown for both states. The blue line denotes the magnon gap. The degrees of freedom below the magnon gap are labelled, including the two zero-energy excitations corresponding to translations (in red). (b) Visualisation of first three eigenvectors of the antiskyrmion and the SP states at $B_z$ = 0.25 T. (c) Lifetime of antiskyrmions in FGT-O (red) at $B_z$ = 0.25 T as a function of inverse temperature (lower $x$-axis) and temperature (upper $x$-axis). For comparison, the lifetime of antiskyrmions in conventional transition-metal ultrathin film Rh/Co/Ir(111) \cite{goerzen2023lifetime} (black) is also shown. The intersection of the two curves and RT is indicated by vertical dotted black lines.}
\end{figure}

Within the HTST framework, we analyze the stability of antiskyrmions by computing the eigenvalue spectrum $\lambda_n$ of the Hessian matrix for both the antiskyrmion and the associated SP in FGT-O as a function of the out-of-plane magnetic field, 
as shown in Fig.~\ref{HTST}(a). From linear spin wave theory, the magnon gap is defined as:
\begin{equation}
\lambda_{\text{mag}} = \mu B_z + 2K~,
\end{equation}
and phenomenologically coincides with the lowest eigenvalue of the FM state.
Eigenmodes with eigenvalues $0 \leq \lambda_n \leq \lambda_{\text{mag}}$ are localized and describe internal degrees of freedom of the soliton, corresponding to low-energy excitations of the spin texture. Their character is identified by visualizing the associated eigenvectors, shown in Fig.~\ref{HTST}(b).

A central and unconventional result is that both the antiskyrmion and the SP exhibit two localized zero modes corresponding to rigid translations (see Appendix D for detailed analysis). In conventional HTST descriptions of soliton annihilation, the SP typically loses at least one translational zero mode. Notably, this conventional behavior is recovered when the isotropic model is employed for FGT-O (see Fig. \ref{eigen_isotropic} in Appendix E). Here, the preservation of the same number of zero modes at the SP reflects a fundamentally different collapse mechanism. This behavior is enforced by symmetry constraints imposed by the aDMI, which promote a collective, spatially extended deformation of the texture along the collapse pathway. Therefore, the decay is governed by a single unstable mode, while surviving translational zero modes at the SP suppress entropic contributions, strongly enhancing the lifetime. In addition, for the antiskyrmion state, the helicity mode disappears because the aDMI breaks the in-plane rotational symmetry of the spin texture and fixes the helicity. By contrast, the breathing mode corresponds to radial deformations of the soliton size and therefore remains a low-energy excitation.

Translations of magnetic configurations describe the collective deflection of the magnetization by some spatial coordinate. We assume that the deflection of magnetic configuration $\mathbf{M}$ by $\mathrm{d}q$ along the translation mode with eigenvector $\mathbf{v}$ can be expressed as a change to the spin-coordinates:
\begin{equation}
\mathbf{M} + \mathrm{d}M = \mathbf{M} + \mathbf{v}\,\mathrm{d}q~.
\end{equation}
For this purpose, we model the magnetic configuration as an interpolated continuous function $\mathbf{M} \to \mathbf{m}(\mathbf{r})$
and write the translation modes action on a single spin
$\mathbf{m}(\mathbf{r}_i)$ as:
\begin{equation}
\mathbf{m}(\mathbf{r}_i) + \mathbf{v}^{\,i}\,\mathrm{d}q
=
\mathbf{m}(\mathbf{r}_i - \mathbf{a}\, \mathrm{d}t)~,
\end{equation}
where $\mathbf{v}^{\,i}$ are the three components of the eigenvector
$\mathbf{v}\in\mathbb{R}^{3N}$, which describe the deflection of the spin $\mathbf{m}(\mathbf{r}_i)$. Also, $t \in [0,1]$ parameterizes the translation of the magnetization by one lattice vector $\mathbf{a}$. 
As a localized object, an isolated soliton spontaneously breaks the continuous translational symmetry of the underlying FM lattice background by existing at a specific coordinate. To restore the symmetry, infinitesimal rigid translations should generate a new configuration with identical energy. This corresponds to the translational zero mode in the Hessian matrix.

For a crystal lattice, its discrete translational symmetry renders its ground state manifold topologically non-trivial. The motion of a defect within such a lattice encounters a periodic restoring force, resulting in an energy barrier known as the Peierls-Nabarro (PN) barrier \cite{hocking2022topological, brizhik2000soliton}. A soliton can be considered as a defect in a periodic magnetic lattice -- a topological defect in a magnetization
field. A similar PN barrier exists for solitons arising from the discreteness of the underlying lattice grid. The height of this barrier depends on the soliton width relative to the lattice constant. A small soliton experiences a large PN barrier, whereas a large soliton
encounters a smaller PN barrier. For a compact spin
texture, which is of the order of the lattice constant, the discrete grid acts as a significant external potential, placing it in a lattice pinned regime. Conversely, for extended textures, the discrete grid samples the magnetization field sufficiently densely, so the lattice is not a significant perturbation. In this limit, the energy landscape for translations is effectively flat. A  flat energy landscape implies zero curvature, and consequently, vanishing eigenvalues $\lambda_n$ of the system's Hessian because:
\begin{equation}
\lambda_n \propto \frac{\partial^2 E}{\partial q_n^2}~.
\end{equation}
In this regime, the texture center can be translated without changing energy, giving rise to two translational zero modes. In the case of FGT-O, the initial states (skyrmion for iDMI and antiskyrmion for aDMI) possess characteristic radii significantly larger than the lattice constant. So, these initial states exhibit well-defined translational zero modes. During the collapse process, the fate of these modes depends on the spatial extent of the SP configuration. If the SP is compact, the soliton center becomes effectively pinned; moving it incurs an energy cost, the instability becomes strongly localized, and the zero modes are lifted. However, if the SP remains extended, the instability does not pin the soliton, and the translational symmetry is preserved. Employing Jacobian transformation from spin- to lattice space \cite{li2026stability} the two translational modes can be represented as: 
\begin{equation} 
\mathbf{v}_x = \partial_x \mathbf{M},~~ \mathbf{v}_y = \partial_y \mathbf{M}~.
\end{equation}
For initial soliton (ini) and extended SP ($\text{e-SP}$), we find:
\begin{equation}
\lambda_x^{\text{i}}=\lambda_y^{\text{i}}=\lambda_x^{\text{e-SP}}=\lambda_y^{\text{e-SP}}=0~. 
\label{eq:zero_ini} 
\end{equation}
Conversely, for compact SP ($\text{c-SP}$):
\begin{equation} 
\lambda_x^{\text{c-SP}} > 0 \;\lor\; \lambda_y^{\text{c-SP}} > 0~. 
\label{eq:lift_trans_compact} 
\end{equation}
Equivalently, the translational derivatives $\partial_x \mathbf{M}_{\text{SP}}$ and $\partial_y \mathbf{M}_{\text{SP}}$ cease to be zero eigenvectors of the Hessian at the SP. As a result, the presence or absence of translational zero modes at the SP provides a sharp
and physically transparent criterion for distinguishing compact and extended SPs. Therefore, we argue that the persistence of translational zero modes constitutes a direct signature
of an extended SP, as enforced by aDMI.


These distinct spectral properties in FGT-O have direct consequences for the pre-exponential factor of HTST. Its evaluation involves only zero modes and positive harmonic modes, yielding the stable-mode contribution
$\sqrt{\prod_{n=1+k_{\text{ini}}}^{2N}\lambda_n^{\text{ini}} \, / \, \prod_{n=2+k_{\text{SP}}}^{2N}\lambda_n^{\text{SP}}}$
which captures the entropic difference between the antiskyrmion and the SP states. Because both states possess the same number of zero modes, apparent by $k_{\text{ini}} = k_{\text{SP}}$, the explicit temperature dependence
$(2\pi k_{\text B} T)^{(k_{\text{ini}}-k_{\text{SP}})/2}$
in the pre-exponential factor [cf.~Eq.~(\ref{pf})] vanishes. This symmetry-driven cancellation plays a key role in determining the thermal lifetime of antiskyrmions in FGT-O.

Fig.~\ref{HTST}(c) shows the resulting antiskyrmion lifetime in FGT-O at $B_z$ = 0.25 T, plotted as a function of inverse temperature (lower $x$-axis) and temperature (upper $x$-axis) on a logarithmic scale. Considering the logarithmic Arrhenius law:
\begin{equation}
\ln \tau = \ln \Gamma_0 + \Delta E/k_{\text{B}}T~,
\label{lifetime_logscale}
\end{equation}
here, the slope of the curve represents the energy barrier, while the offset at $T^{-1} \rightarrow 0$ represents $\Gamma_0$. For comparison, we also include the lifetime of antiskyrmions in the prototypical ultrathin Rh/Co/Ir(111) film \cite{goerzen2023lifetime}. Remarkably, in the low-temperature regime, the lifetime of antiskyrmions in FGT-O is comparable to that reported for Rh/Co/Ir(111), while in the high-temperature regime (near RT), it exceeds the latter by up to 5 orders of magnitude. At 300 K, the lifetime exceeds 0.2 ns. This pronounced enhancement reflects the modified energy landscape induced by the aDMI, which strongly suppresses thermally activated collapse at elevated temperatures. Consequently, nanoscale antiskyrmions in FGT-O are expected to be sufficiently long-lived to enable direct experimental detection and controlled manipulation using the state-of-the-art techniques, such as spin-polarized scanning tunneling microscopy \cite{muckel2021experimental} and Lorentz transmission electron microscopy \cite{nayak2017magnetic,yalisove2026mapping}.

Previous studies have shown that the pre-exponential factor can fundamentally limit lifetimes through entropic effects \cite{goerzen2023lifetime}. In particular, Wild \textit{et al.} \cite{wild2017entropy} demonstrated that the exponential proliferation of decay pathways leads to extreme enthalpy-entropy compensation, causing the pre-exponential factor to vary over 30 orders of magnitude and dramatically reducing skyrmion lifetimes at elevated temperatures. Here, we demonstrate the opposite limit. aDMI breaks rotational symmetry and enforces a collective, spatially extended SP. The extended transition state does not pin the soliton position, allowing translational zero modes to survive at the SP. As a consequence, the entropic contribution to the decay of the pre-exponential factor is suppressed, resulting in nearly temperature-independent lifetimes. Our work shows that aDMI does not merely modify the energy barrier but qualitatively reshapes the geometry of the transition state. A temperature-independent prefactor yields
more predictable and stable lifetimes across a broader range of physical parameters. This leads to a new engineering approach to achieve long lifetimes by exploiting aDMI-induced symmetry breaking to control the entropic effect, rather than relying solely on increasing energy barriers.

\section{CONCLUSIONS AND OUTLOOK}

In summary, we develop a first-principles method for computing aDMI in triangular lattices that goes beyond the isotropic approximation commonly used to date. Using an experimentally feasible system, monolayer FGT-O, in which in-plane symmetry is broken, as a representative example, we demonstrate the decisive role of aDMI in stabilizing chiral magnetic textures. We find that aDMI stabilizes antiskyrmions instead of skyrmions, as reported in other materials \cite{hoffmann2017antiskyrmions,niu2025magnetic}. Remarkably, we reveal a new lifetime regime in which antiskyrmions exhibit strongly temperature-independent stability. To the best of our knowledge, this constitutes the first report of such a lifetime regime in chiral magnetic textures. The microscopic origin of this behavior is from the symmetry breaking enforced by aDMI, which not only selects antiskyrmions as the energetically favored solitons but also prevents a compact collapse, leading to spatially extended SP spin textures. As a result, both the metastable antiskyrmion and the SP share the same fluctuation spectrum, with two translational zero modes surviving in both states. Because the zero mode structure is identical in both states, their contributions cancel in the prefactor, suppressing entropic enhancement of the decay rate. More generally, we identify the conditions under which spatially extended SPs arise in soliton collapse. These results establish general design principles for engineering soliton stability by controlling transition-state geometry, enabling decay pathways with weakly temperature-dependent lifetimes.

\begin{figure}[t]
	\centering
	\includegraphics[width=1.0\linewidth]{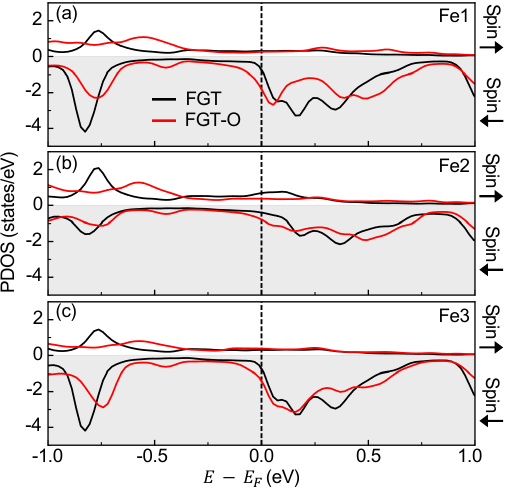}
	\caption{\label{pdos} Spin-resolved PDOS on (a) Fe1, (b) Fe2, and (c) Fe3 atoms for FGT (black) and FGT-O (red). Here, Fe1 directly bonds to the oxygen atom.
    Fermi energy ($E_{\text{F}}$) is set to zero. The spin-up and spin-down are shown in positive and negative $y$-axes, respectively. }
\end{figure}%

\begin{figure}[t]
	\centering
	\includegraphics[width=0.8\linewidth]{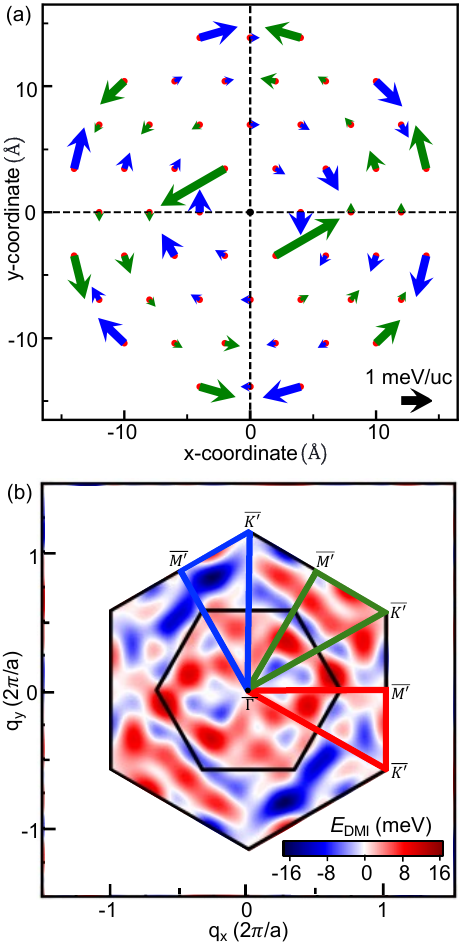}
	\caption{\label{aDMI_analysis}
    (a) Shell-resolved in-plane aDMI constants for FGT-O up to 7$^\text{th}$ NN shown in real space. The arrow length depicts the magnitude of aDMI constants, with a black reference arrow (bottom right) indicating the scale. For clarity, vectors with magnitudes below 0.3 meV/uc were uniformly rescaled while preserving their directions; 0.3 meV/uc thus defines the visual cutoff. Blue and green colors denote clockwise and counterclockwise rotations, respectively.
    (b) aDMI energy map in reciprocal space. The three high-symmetry paths used in the analysis are indicated in red, green, and blue.}
\end{figure}%

\begin{table*}[htbp]
\caption{\label{table1} Shell-resolved exchange constants ($J_i$) and iDMI constants ($D_i$) for FGT-O up to 8$^\text{th}$ and 7$^\text{th}$ NN, respectively. A positive (negative) sign of $J_i$ represents FM (AFM) coupling. A positive (negative) sign of $D_i$ denotes a preference for clockwise (counter-clockwise) rotating cycloidal spin spirals. All values are given in meV/uc.}
\renewcommand{\arraystretch}{1.4}
\begin{ruledtabular}
\begin{tabular}{ccccccccccccccccc}
Interactions (meV/uc)&\textit{$J_1$}&\textit{$J_2$}&\textit{$J
_3$}&\textit{$J_4$}&\textit{$J_5$}&\textit{$J_6$}&\textit{$J_7$}&\textit{$J_8$}&\textbf{$D_1$}&\textbf{$D_2$}&\textbf{$D_3$}&\textbf{$D_4$}&\textbf{$D_5$}&\textbf{$D_6$}&\textbf{$D_7$}\\
\hline
Path 1 (red) &11.30&12.66&4.72&-2.72&-1.64&-0.82&0.43&1.32&-0.39&-0.88&0.33&0.29&0.03&0.06&-0.19\\
\hline
Path 2 (green) &15.84&7.89&6.04&-2.52&-2.68&1.79&0.38&0.28&-0.80&0.25&0.12&0.77&-0.55&-1.48&0.78\\
\hline
Path 3 (blue) &10.59&11.75&3.98&-0.58&-2.89&0.07&-0.77&1.13&-0.09&-0.10&-0.83&0.29&-0.45&-0.44&0.19\\
\end{tabular}
\end{ruledtabular}
\end{table*}

\begin{figure}[t]
	\centering
	\includegraphics[width=1.0\linewidth]{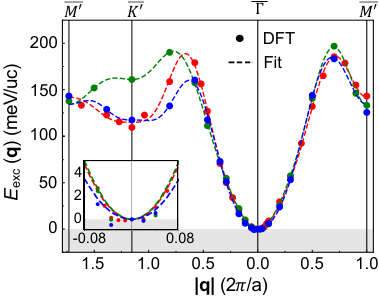}
	\caption{\label{Exchange_fit} Flat spin spiral energy dispersion of FGT-O along $\overline{\text{M}}$-$\overline{\Gamma}$-$\overline{\text{K}}$-$\overline{\text{M}}$ without SOC. Symbols represent scalar relativistic DFT results, while dashed and solid lines show Heisenberg model fits including up to seventh NNs. Dispersions along the 3 high symmetry directions correspond to those in Fig. \ref{Material_RMSD_SpinSpiral}. The zoom in near the $\Gamma$ point is shown as an inset.}
\end{figure}%

Antiskyrmions are more promising than skyrmions for topological Hall physics. Although both carry a finite topological charge of $|Q|=1$, their transport responses are governed by different symmetry properties. The rotational symmetry of skyrmions leads to an isotropic emergent magnetic field arising from scalar spin chirality, $\chi_{ijk} = \mathbf{m}_i \cdot \left( \mathbf{m}_j \times \mathbf{m}_k \right)$, and hence a largely fixed Hall response. Antiskyrmions host anisotropic spin chiralities which produce a symmetry-encoded, direction-dependent emergent magnetic field that allows the transverse response to be selectively enhanced or suppressed \cite{nayak2017magnetic,Kumar2020}. In addition, antiskyrmions give rise to enhanced topological orbital moments even without SOC compared to skyrmions, consistent with chirality-driven orbital responses \cite{Gobel_2019_PRB}, opening a route toward \textit{topological orbitronics} \cite{oh2025interplay}.

Beyond this, recent experiments \cite{sharma2024gate} have demonstrated that bilayer FGT-O hosts an AFM ground state, positioning this system as a highly promising platform for synthetic AFM solitons. In such a system, a synthetic soliton consists of two FM antiskyrmions hosted in separate layers and coupled antiferromagnetically. These textures are expected to combine the intrinsic stability induced by aDMI with the dynamical advantages of AFM solitons, including strongly enhanced current-driven velocities and suppressed transverse motion \cite{pham2024fast}. The coexistence of aDMI, antiskyrmion stability, and AFM interlayer coupling positions bilayer FGT-O as an experimentally accessible and attractive platform for future high-speed soliton-based spintronic devices. Our work establishes aDMI as a new avenue for stabilizing chiral magnetic textures beyond conventional barrier engineering.

\section*{APPENDIX A: PROJECTED DENSITY OF STATES}
\label{pdos_analysis}

We analyzed the PDOS of the inequivalent Fe atoms to understand how oxygen adsorption modifies the electronic structure near \(E_{\mathrm F}\), as shown in Fig.~\ref{pdos}. Upon oxygen adsorption, the Fe states exhibit a pronounced redistribution of spectral weight around \(E_{\mathrm F}\), characterized by a suppression of spin-up states and a concomitant enhancement of spin-down states, indicating strong Fe--O hybridization.

The effect is most pronounced for the Fe1 atom, which is directly bonded to the oxygen atom. In this case, a prominent peak located just above \(E_{\mathrm F}\) in pristine FGT shifts significantly and changes its spectral weight after oxygen adsorption, reflecting the strong hybridization between Fe1-$3d$ states and O-$2p$ orbitals. In contrast, the electronic states of Fe2 and Fe3 are modified more moderately, consistent with their larger distance from the adsorbed oxygen atom and the more indirect nature of their hybridization.

\begin{figure}[b]
	\centering
	\includegraphics[width=0.9\linewidth]{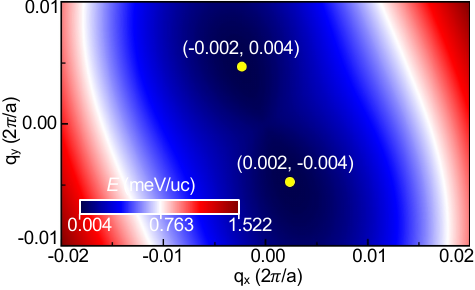}
	\caption{\label{rad_barr_minima} Energy landscape in $\mathbf{q}$ space near the $\Gamma$ point, revealing two symmetry-related spin-spiral minima at finite wave vector coordinates (-0.002, 0.004) and (0.002, -0.004) shown by yellow dots, indicating that the ground state favors spin spirals.}
\end{figure}%

Despite this substantial local reconstruction of the electronic structure, the total spin moment of FGT-O changes only slightly, indicating that the hybridization mainly redistributes spin-resolved spectral weight without significantly altering the overall magnetic moments.

\section*{APPENDIX B: INTERACTION PARAMETERS}
\label{parameters}

Fig.~\ref{Exchange_fit} depicts energy dispersions of flat spin spirals for the FGT-O monolayer along the three high-symmetry paths, computed without SOC under isotropic model conditions. The DFT energies along each path (red, green, and blue) were fitted separately to the exchange Hamiltonian in order to extract the corresponding exchange parameters. Table~\ref{table1} summarizes the Heisenberg exchange and the DMI constants obtained within the isotropic model applied to the three high-symmetry paths. Since three Fe atoms within a unit cell are treated collectively, the extracted parameters correspond to an effective 2D spin model. We employ monolayer fitting functions, approximating different layers in our system to be equivalent to an effective single layer. The interlayer interaction contributions are thus implicitly incorporated within $J_n$/$D_n$. For the anisotropic model, the exchange constants are taken as the average of the isotropic exchange parameters obtained along the three paths, while the full set of aDMI constants is shown in Fig.~\ref{aDMI_analysis}(a). The corresponding aDMI energy landscape in reciprocal space, computed from these parameters, is presented in Fig.~\ref{aDMI_analysis}(b). The contour map reveals pronounced directional variations, directly reflecting the crystallographic anisotropy of the DMI.

\section*{APPENDIX C: GROUND STATE AT ZERO FIELD}
\label{groundstate}

Fig.~\ref{rad_barr_minima} presents the reciprocal space energy dispersion of spin spirals for FGT-O. The contour map depicts the energy variation defined as the total exchange and aDMI energies of a spin spiral relative to the FM state, with an additional contribution of half the MAE, $E(\mathbf{q})=E_{\text{exc}}(\mathbf{q})+\Delta E_{\text{SOC}}(\V{q})+K/2$. The identified minima of the dispersion lie extremely close to the FM state in reciprocal space and are nearly degenerate in energy with it.

\begin{figure}[t]
	\centering
	\includegraphics[width=1.0\linewidth]{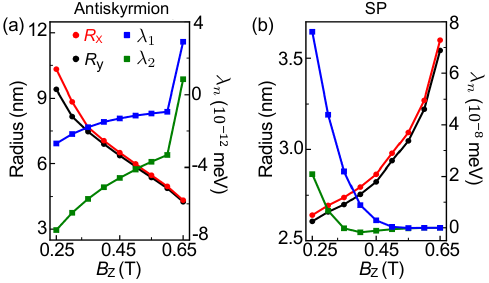}
	\caption{(a) Radius (left $y$-axis) and the first two non-negative eigenvalues (right $y$-axis) as functions of $B_z$ for the antiskyrmion state. (b) Same as (a), but for the SP state.
}
\label{eigen_anisotropic}
\end{figure}%

\section*{APPENDIX D: EIGENVALUE ANALYSIS IN THE ANISOTROPIC MODEL}
\label{eigen_aniso}

We analyze the evolution of the lowest two non-negative eigenvalues of the Hessian and the corresponding soliton radius as functions of the magnetic field for both the metastable antiskyrmion and the SP along the collapse pathway, as shown in Fig.~\ref{eigen_anisotropic}. The soliton radius (left $y$-axis) and the first two non-negative eigenvalues (right $y$-axis) are plotted as functions of $B_z$. As $B_z$ increases, the radius of the metastable antiskyrmion decreases, whereas the SP radius increases slightly. When the field approaches the critical value $B_z = 0.65$ T, the two radii converge, indicating the disappearance of the activation barrier ($\Delta E = 0$) and the onset of the instability of the antiskyrmion state.

For the SP state, the eigenvalues shown on the right $Y$ axis decrease in magnitude with increasing $B_z$, reflecting the progressive flattening of the energy landscape as the system approaches the critical field. At low $B_z$, the weak background field produces a steep energy landscape with large curvature, resulting in relatively large eigenvalues. As a consequence, large-scale deformations of the antiskyrmion are energetically costly. As $B_z$ increases and the energy barrier is reduced, the curvature of the energy landscape along the relevant deformation modes decreases, leading to smaller eigenvalues. This trend is consistent with the energy tuning mechanism illustrated in Fig.~\ref{SP_transition}, where increasing the magnetic field shifts the SP toward the extended regime, resulting in a more extended SP configuration and a flatter energy landscape. Close to the critical field, the instability becomes spatially more extended and involves collective fluctuations of the antiskyrmion profile, consistent with the extended character of the SP configuration induced by the aDMI.

\begin{figure}[b]
	\centering
	\includegraphics[width=1.0\linewidth]{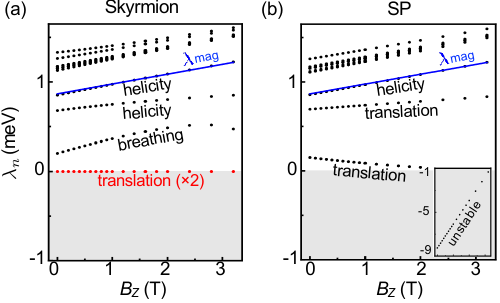}
	\caption{Hessian eigenvalue spectra for the skyrmion and the SP in the isotropic model.
(a) Skyrmion state: two translational zero modes (red) persist at $\lambda_n \approx 0$. The blue line denotes
the magnon gap.
(b) SP state along the skyrmion to FM collapse path: the translational modes are lifted, and a single unstable mode emerges (gray shaded region, see inset), as expected for a localized transition state.}
\label{eigen_isotropic}
\end{figure}%

Note that in our analysis, eigenvalues within the interval $[-10^{-4}, 10^{-4}]$ meV are treated as zero modes, corresponding to degrees of freedom associated with continuous symmetries of the system. Throughout the entire magnetic-field range considered here, both the metastable antiskyrmion and the SP exhibit two such translational zero modes as discussed in the main text.

\section*{APPENDIX E: EIGENVALUE SPECTRUM IN THE ISOTROPIC MODEL}
\label{eigen_iso}

In Fig.~\ref{eigen_isotropic}, we plot the curvatures, given by the eigenvalues $\lambda_n$ of the spherical-constraint Hessian for the isotropic model. For the skyrmion state, two translational zero modes ($\lambda_n \approx 0$) appear, while higher eigenmodes correspond to breathing, helicity, and magnon excitations. Here, in contrast to the aDMI case, helicity modes remain because iDMI preserves the in-plane rotational symmetry. At the SP, these translational zero modes are lifted, resulting in a finite entropic contribution to the decay prefactor and giving rise to conventional Arrhenius behavior in the isotropic model. This is completely different from the zero modes at the SP in the anisotropic model.

\section*{ACKNOWLEDGMENTS}
	
This study has been supported through the ANR Grant No. ANR-22-CE24-0019. This work is supported by France 2030 government investment plan managed by the French National Research Agency under grant reference PEPR SPIN – [SPINTHEORY] ANR-22-EXSP-0009. This study has been (partially) supported through the grant NanoX no.~ANR-17-EURE-0009 in the framework of the ``Programme des Investissements d’Avenir". This work was performed using HPC resources from CALMIP (Grant No. 2024/2026-[P21023]). 
	

\emph{}

The authors declare no competing interests.

\section*{DATA AVAILABILITY}
The data that support the findings of this article are not publicly available. The data underlying this study is available from the authors upon reasonable request.

\bibliography{References}

\end{document}